\documentclass[amsmath,amssymb,twocolumn]{revtex4-2}
\usepackage{graphicx}
\usepackage{subfig}
\usepackage{epsfig}
\usepackage{dcolumn}
\usepackage{multirow}
\usepackage{rotating}
\usepackage{verbatim}
\usepackage{bm}
\usepackage{color}
\usepackage{soul}
\usepackage{float}
\usepackage{hyperref}
\usepackage[utf8]{inputenc}

\def\lsim{\raise0.3ex\hbox{$<$\kern-0.75em\raise-1.1ex\hbox{$\sim$}}}
\def\gsim{\raise0.3ex\hbox{$>$\kern-0.75em\raise-1.1ex\hbox{$\sim$}}}

\def\beqa{\begin{eqnarray}}
\def\eeqa{\end{eqnarray}}

\begin{document}

\title{Study of the azimuthal asymmetry in heavy ion collisions combining initial state momentum orientation and final state collective effects}
\author{L.S. Moriggi$^{1}$}
\email{lucasmoriggi@unicentro.br}
\affiliation{$^{1}$ Universidade Estadual do Centro-Oeste (UNICENTRO), Campus Cedeteg, Guarapuava 85015-430, Brazil}
\author{\'E.S. Rocha$^{2}$} 
\email{erison.rocha@ufrgs.br}
\author{M.V.T. Machado$^{2}$}
\email{magnus@if.ufrgs.br}
\affiliation{$^{2}$ High Energy Physics Phenomenology Group, GFPAE. Institute of Physics, Federal University of Rio Grande do Sul (UFRGS)\\
Caixa Postal 15051, CEP 91501-970, Porto Alegre, RS, Brazil}

\begin{abstract}
In the present work we investigate the source of azimuthal asymmetry for nuclear collision using a model that contemplates particles produced in the initial hard collisions and the collective effects described by a Blast-Wave like expansion. The latter is described by the relaxation time approximation  of the Boltzmann transport equation.  The parameters regarding collective flow and asymmetry are fitted by the experimental data from $p_T$ spectrum and $v_2$ for PbPb and XeXe collisions at different centrality classes. As a by-product the ratio of final elliptic flow with the initial anisotropy, $v_2/\epsilon_2$, and the average transverse momentum are predicted.

\end{abstract}

\maketitle

\section{Introduction}
The available relativistic heavy ion collision experiments allow us to investigate the underlying dynamics of quarks and gluons at very high energy density \cite{Arslandok:2023utm}. The thermalized deconfined parton system, i.e. the Quark-Gluon Plasma (QGP) \cite{Shuryak:1978ij,Shuryak:1980tp}, generated in these reactions has important signatures like the collective flow, parton energy loss, quarkonium production suppression and many others \cite{Das:2022lqh,Ritter:2014uca}. In particular, the elliptic flow \cite{Ollitrault:1992bk,Voloshin:2009fd,Voloshin:1994mz,Poskanzer:1998yz} ($v_2$, the second harmonic coefficient of the azimuthal Fourier decomposition of the momentum distribution) measures the non-uniformity of the flow in all directions as viewed along the beam-line \cite{Snellings:2011sz,Heinz:2013th,MadanAggarwal:2021,Ollitrault:2023wjk}. Namely, it characterizes the azimuthal momentum space anisotropy of particle emission from non-central heavy-ion collisions in the plane transverse to the beam direction. As the anisotropy is largest in the first instance of the system evolution, the $v_2$-coefficient is quite sensible to the early stages of the collisions. Thus, elliptic flow encodes the residual asymmetry of the particle density in momentum space referring to the reaction plane subsequent to the hadronization. 

The main motivation of present work is to address the simultaneous description of the nuclear modification factor, $R_{AA}$, and the elliptic flow in a consistent way.  It is already known that parton energy loss models that have been tuned to describe $R_{AA}$ in identified hadron production underestimate the elliptic flow $v_2$ at intermediate transverse momentum \cite{Molnar:2013eqa,Noronha-Hostler:2016eow,Zhang:2013oca,Liao:2008dk,Kopeliovich:2012sc,Cao:2017umt,Andres:2019eus}. This has been named in literature as the $R_{AA}\otimes v_2$ puzzle.  Both observables are nicely described by the QGP´s hydrodynamic expansion as a strongly coupled fluid  at low transverse momentum ($p_T \lesssim 2$ GeV) \cite{Romatschke:2007mq,Song:2010mg,Gale:2012rq,Huovinen:2013wma,Heinz:2013th,Gale:2013da,Niemi:2015qia,Bernhard:2016tnd,McDonald:2016vlt,Zhao:2017yhj}. At high momentum, typically  $p_T\gtrsim 10$ GeV, they can be correctly described in terms of jet quenching due to hard parton energy loss in their propagation across the hot QGP
medium \cite{Wang:1992qdg,Wang:1998ww,Vitev:2002pf,Wang:2002ri,Eskola:2004cr,Qin:2007rn,Schenke:2009gb,Chen:2011vt,Majumder:2011uk,Buzzatti:2011vt,Zapp:2012ak,Gyulassy:2000gk,Kumar:2017des,Zigic:2019sth,Mehtar-Tani:2013pia,Qin:2015srf,Blaizot:2015lma}. On the other hand, it is a challenge to describe $R_{AA}$ and $v_2$ consistently in the intermediate transverse momentum region characterized by the soft and hard physics confluence. In our case, the parton/gluon saturation physics embedded into the theoretical approach allows to use weak coupling methods in this interface region. Recently, studies using a combination of the state-of-art on transport models, hadronization and hydrodynamic evolution  have been carried out. For instance, in Ref.  \cite{Zhao:2021vmu} both quark coalescence and a hadronic afterburner in the COLBT-hydro model are implemented. That investigation combines event-by-event hydrodynamics, jet quenching as well as hadron cascade and simultaneously describe $R_{AA}$ and $v_2$ in the full range of transverse momentum. Similar approaches have been proposed, see \cite{Barreto:2022ulg,Stojku:2020wkh},  which will shed light into further investigations of the topic. In present study the question is to what
extent the elliptic flow can be determined by the initial state effects and how these effects can be detached from the final state ones.

Concerning identified particle spectra in transverse momenta ($p_T$), the hadron production can be described within the  $k_T$-factorization formalism (including the primordial parton transverse momenta) and one considers that the cold matter nuclear  effects are generated in the hard interaction of the nuclei at initial states of the corresponding collision. On the other hand,  afterwards those systems undergo a hydrodynamics evolution to freeze-out   which alters the corresponding  $p_T$-spectra. In the context of the relaxation time approximation (RTA) of the  Boltzmann transport equation (BTE) \cite{Florkowski:2016qig}    the $p_T$ spectrum can be described by performing a temporal separation  between  hadrons produced in initial state hard collision and those produced in the equilibrium situation \cite{Tripathy_2016,Tripathy:2017kwb,Qiao_2020}. In Refs. \cite{Moriggi:2020qla,Moriggi:2022xbg} we considered this approach to describe the spectra of light hadrons in lead-lead (PbPb) collisions at the Large Hadron Collider (LHC) and  at the Relativistic Heavy Ion Collider (RHIC) as well. In particular, the approach has been used to describe particle production in small systems at RHIC \cite{Moriggi:2022xbg}.  The nuclear modification factors for pAl, pAu, dAu and HeAu have been successfully reproduced as a function of $p_T$ for different centralities. An important result was that the thermal parametrization can be considerably modified by taking into consideration the nuclear effects embedded in the gluon distribution function of the target. 

In this work we investigate the possible sources of azimuthal asymmetry for relativistic heavy ion collisions using a theoretical model that contemplates particles produced in the initial hard collisions and the collective effects described by a Blast-Wave like expansion.  The interface between the hard process described within the QCD $k_T$-factorization formalism and the final state collective effects is studied in details. In the  hard part of spectrum a contribution to the elliptic flow is included associated  to the azimuthal orientation of the momentum of the produced particles in relation to the reaction plane. This effect is introduced in  the QCD color dipole amplitude from which the unintegrated gluon distribution is obtained. The charged hadron production in proton-proton cross section has been described successfully using this approach in Ref. \cite{Moriggi:2020zbv}. In addition, azimuthal anisotropy is also incorporated in the BTE within the RTA approximation. There, the transverse rapidity variable $\rho$ has been modified in order to include an anisotropy dependence in the flow. In this last context we are following closely the Refs. \cite{Tripathy:2017nmo,Younus:2018mrk,Huovinen:2001cy,Akhil:2023xpb}. The nuclear modification factor $R_{AA}$ and $v_2$ are described simultaneously for PbPb and XeXe collisions at the LHC. 

This paper is organized as follows. In Sec. \ref{sec:theory} the theoretical model for the charged hadron production is presented in the context of the $k_T$-factorization approach including azimuthal asymmetry and nuclear shadowing. This initial distribution is then embedded in the formalism of hydrodynamical blast wave model (BTE-RTA approach).  Expressions for both the nuclear modification factor,  $R_{AA}$, and elliptic flow, $v_2$ are provided.  In Sec.  \ref{sec:results} the comparison of the results to the LHC data are done and the interplay between initial and collective azimuthal asymmetries is investigated.  Finally, we summarize our conclusions in the last section  \ref{sec:conc}.

\section{Theoretical framework and main predictions}
\label{sec:theory}

We shall study the anisotropic flow, which is one of the key observables to analyse the transport properties of the QGP. It is determined by the flow harmonic coefficients $v_n$ obtained from the Fourier decomposition of the azimuthal distribution of the produced
particles in the following way \cite{Ollitrault:1992bk,Voloshin:2009fd,Voloshin:1994mz,Poskanzer:1998yz}:
\begin{equation}
E\frac{d^3N}{d^3\vec{p}} =\frac{d^2N}{2\pi p_Tdp_Tdy}\left( 1 + 2 \sum\limits_{n=1}^{\infty}v_{n}\mathrm{cos}(n\phi)\right),
\label{eq: def}
\end{equation}
where $\phi$ is the azimuthal angle with respect to the
reaction plane. Accordingly, $E$ is the particle energy with $p_T$  and $y$ being its transverse momentum value and rapidity.

The corresponding anisotropy for an identified hadron $h$ in a inclusive nucleus-nucleus collision, $AA\rightarrow hX$, is characterized by the set of Fourier flow coefficients $v_n$, defined as:
\begin{eqnarray}
v_n(p_T) \equiv \frac{\int_0^{2\pi} d\phi \cos(n\phi) \frac{d^3\sigma (AA\rightarrow hX)}{dy d^2\vec{p}_T}}{\int_0^{2\pi} d\phi \frac{d^3\sigma (AA\rightarrow hX)}{dy d^2\vec{p}_T}}    ,
\end{eqnarray}
with $y$ and $\vec{p}_T$ being the rapidity and transverse momentum vector of the produced hadron, respectively. The integration is done over the azimuthal angle $\phi$ in coordinate space.

Here, the RTA approximation of the BTE will be considered which is an effective model where the collisional term has the form $\mathrm{C}[f] = -(f-f_{eq})/t_r$. The Boltzmann local equilibrium distribution, $f_{eq}$, for the distribution of particles $f$ is typified by a freeze-out temperature $T_{eq}$ and $t_r$ is the relaxation time. The latter corresponds to the time for a non-equilibrium system to reach the equilibrium whereas $t_f$ is the freeze-out time parameter. Given $\mathrm{C}[f]$, the BTE is then solved with the following initial conditions, $f(t=0)=f_{in}$ and $f(t=t_f)=f_{fin}$. Therefore, the final distribution $f_{fin}$ is evaluated as a function of the ratio $t_f/t_r$ \cite{Tripathy:2017kwb},
\begin{eqnarray}
f_{fin}=f_{eq}+\left(f_{in}-f_{eq}  \right)e^{-\frac{t_f}{t_r}},    
\label{eq:finexp}
\end{eqnarray}
with $f_{fin}\rightarrow f_{eq}$ if the system is given enough time or $t_f>t_r$. 

Let us now focus on the determination of the initial distribution of particles in the context of the high energy factorization approach. In this case, the inclusive cross section for producing identified  particles is given in terms of the convolution of the unintegrated gluon distribution (UGD) for both target and projectile and the gluon-gluon sub-process cross section.  The distribution $f_{in}$ is proportional to the production multiplicity,  $d^3N (AA\rightarrow hX)/dy d^2\vec{p}_T$. The UGDs containing initial state nuclear effects can be computed within the QCD color dipole framework. The main advantage is that the input dipole-target scattering amplitude can be constrained from experimental data in the small-$x$ regime.  The dipole-nuclei $S$-matrix at transverse distance $r$ coordinate space for a given impact parameter $b$ can be computed from the   dipole-proton cross-section, $\sigma_{p}(x,r)$, considering the Glauber-Mueller multiple-scattering formalism \cite{Glauber:1955qq,Mueller:1989st},
\begin{eqnarray}
\label{eq:SdA}
S_{A}(x,r,b)=e^{-\frac{1}{2} T_A(b)\sigma_{p}(x,r)},   
\end{eqnarray}
where $T_A(b)$ is the nuclear thickness function and $b$ is the distance to the nuclei center. The nuclear $S$-matrix is a function of the gluon longitudinal momentum fraction, $x$, the transverse size of the QCD color dipole $r$ and $b$. We are considering the proton as a homogeneous target in the impact parameter, with transverse area $\pi R_p^2=\sigma_0/2$, so that $\sigma_{p}(x,r)=2\int d^2b (1-S_p(x,r,b))=\sigma_0(1-S_p(x,r))$. The last quantity is the dipole-proton cross section.
The scattering matrix for the proton case, $S_p$,  can be obtained from the Fourier transform of the UGD as a function of transverse momentum. In this work we are considering the UGD parametrization proposed by Moriggi, Peccini, and Machado fitted to the $p_T$ spectrum at different energies in proton-proton collisions \cite{Moriggi:2020zbv}. The corresponding analytical dipole-proton cross section is given by,
\begin{eqnarray}
\label{eq:DISc}
\sigma_{p}(\tau_r)=\sigma_0\left (  1-\frac{2(\frac{\tau_r}{2})^{\xi}K_{\xi}(\tau_r)}{\Gamma(\xi)} \right ),
\end{eqnarray}
where $\tau_r=Q_s(x)r$ is the scaling variable as a function of $r$ and $K_\xi(\tau_r)$ the Bessel function of the second kind. The saturation scale and the power parameter $\xi$ are fitted considering the geometric scaling in the momentum spectrum $\tau_Q=Q^2/Q_s^2(x)$, so that $Q^2=p_T^2$,
\begin{eqnarray}
\label{eq:pars}
 \xi &=& 1 + a \tau_Q ^b ,\\
 Q_s^2(x)&=& \left( \frac{x_0}{x}\right) ^{0.33}.
\end{eqnarray}
The parameters $a$, $b$, and $x_0$ are given by Moriggi et al in Ref.  \cite{Moriggi:2020zbv}. The main nuclear effect in Eq. \eqref{eq:SdA} is to introduce shadowing and the Cronin peak\cite{Moriggi:2020qla}. For $r\rightarrow 0$ the QCD dipole-nucleus cross section is $\sigma_A (x,r)\approx A\sigma_p(x,r)$, which implies in a scaling of binary collisions $N_{coll}$ for $p_T\gg Q_s(x)$ and consequently the nuclear modification factor $R_{AA} \rightarrow 1$.

\begin{figure*}[t]
\includegraphics[width=0.7\linewidth]{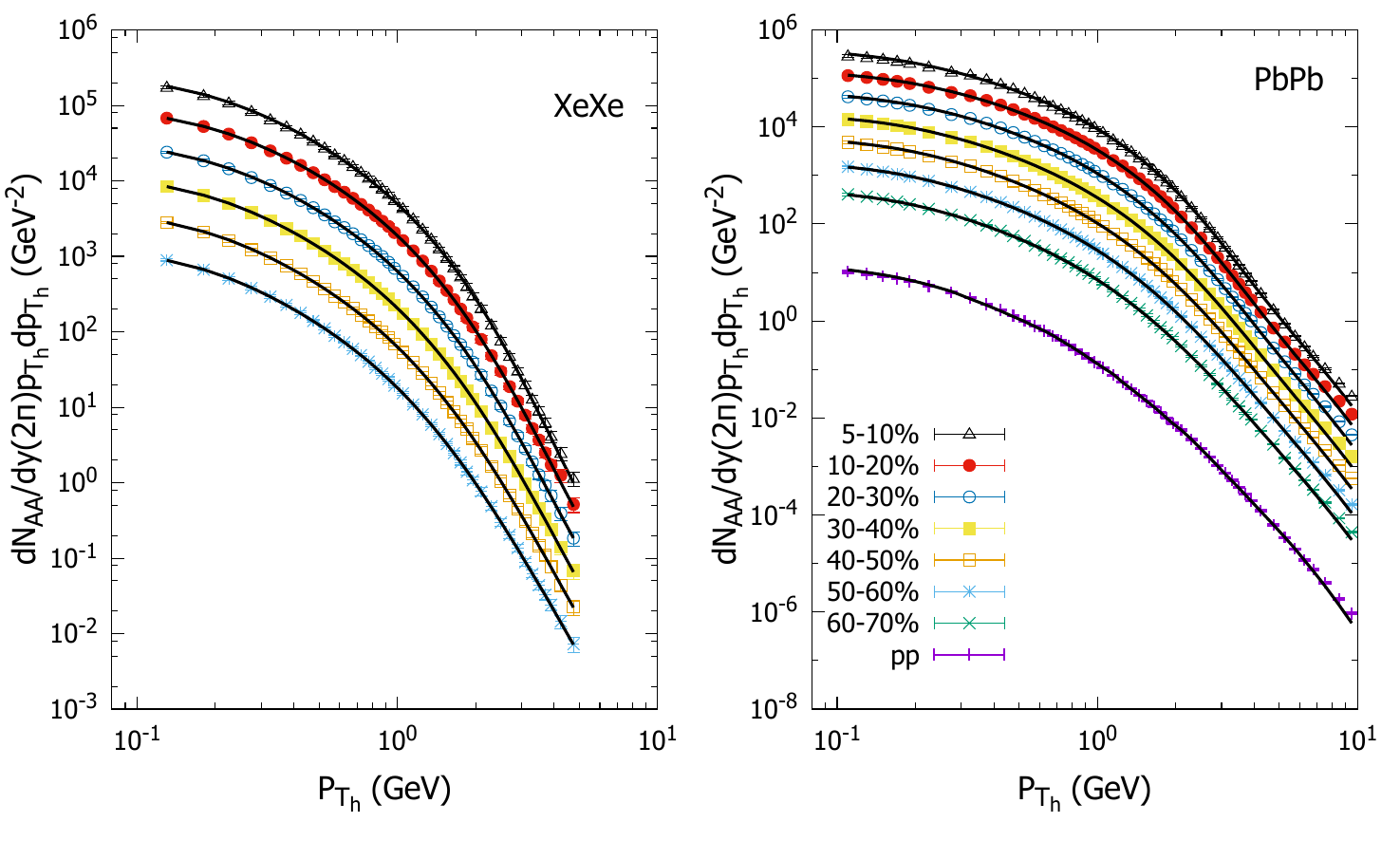}
\caption{Multiplicity of $\pi^\pm$ produced in PbPb and XeXe collisions at $\sqrt{s}=5.02 $ TeV and $\sqrt{s}=5.44$ TeV, respectively, compared with experimental data from ALICE collaboration \cite{ALICE:2019hno,ALICE:2021lsv} for different centrality classes. Each curve has been multiplied by a factor of $2^i$ for a  better displaying.}
\label{fig:NAA} 
\end{figure*}

An azimuthal asymmetry can be generated if there is a dipole orientation described by the transverse size $\vec{r}$ regarding the impact parameter $\vec{b}$ (the angle between them is denoted by $\phi_{rb}$). Namely, the scattering amplitude will depend upon the angle between $\vec{r}$ and $\vec{b}$ which should  introduce a dependence on angle $\phi_p$ between $\vec{p}_T$ and $\vec{B}$ (the impact parameter of the nucleus-nucleus collision). The evaluation of the amplitude taking into account the dipole orientation is  not an easy task (see Refs. \cite{Iancu:2017fzn,Kopeliovich:2021dgx,ALTINOLUK2016373,PhysRevD.99.074004,PhysRevD.100.034007} for more detailed theoretical analyzes). The numerical solution of the Balitsky-Kovchegov (BK) equation including dipole orientation has been studied in Ref. \cite{Golec-Biernat:2003naj,PhysRevD.83.034015}. In what follows we consider a way to introduce such asymmetry in a more fundamental level as proposed by Iancu and Rezaeian \cite{Iancu:2017fzn}. It was shown that by taking into account a non homogeneous color charge sources distribution for the target in the context of the Color Glass condensate Condensate (CGC) formalism one produces $S_A = \exp[-{\cal{N}}_{2g}]$ with ${\cal{N}}_{2g}(r,b,\phi_{rb}) = {\cal{N}}_0(r,b)+{\cal{N}}_{\theta}(r,b)\cos(2\phi_{rb})$. Expressions are provided in \cite{Iancu:2017fzn} for the auxiliary functions ${\cal{N}}_{0,\theta}$ for the case of a Gaussian distribution.

In this paper, we chose a simpler way to incorporate this effect from a phenomenological perspective, by angular modulation, given by the substitution
\begin{equation}
    r^2 \rightarrow r^2\left[1+a_r\cos(2\phi_{rb})\right].
\end{equation}
The parameter $a_r$ will be fitted from the experimental data of each centrality class, and it measures the amount of asymmetry needed to describe $v_2$. Such modification should give a non-zero $v_2$ coefficient in the momentum spectrum of produced gluons. The cross section for inclusive gluon production with transverse momentum $p_T$ in the $k_T$-factorization formalism, can be described with the dipole scattering matrix $S_{A}(x,r,b)$ in the position space and the corresponding initial particle distribution is given by:

\begin{eqnarray}\label{eq:fatkt}
f_{in}(\phi_p) &=& \frac{1}{p_T^2}\frac{2C_F}{(2\pi)^4\alpha_s} \int d^2b d^2r e^{i \vec p_T\cdot \vec r} \nabla_r^2 S_A(x_1,r,b) \nonumber \\ 
&\times& \nabla_r^2 S_A(x_2,r,b'),
\end{eqnarray}
such that $\vec b'=\vec b-\vec B$ and $x_{1,2}=p_T e^{\pm y}/\sqrt{s}$ are the momentum fractions carried by the gluons for each nuclei and $\nabla_r^2$ is the Laplacian with respect to $r$ coordinate. We can decompose the distribution $p_T$ into its harmonic components using the identity
\begin{equation}
e^{ip_Tr \cos(\phi_p-\phi_r)}=\sum_{n=-\infty }^{n=\infty } i^n J_n(p_Tr)e^{in(\phi_p - \phi_r)},
\end{equation}
where $J_n(x)$ is the Bessel function of first kind of order $n$ and $\phi_r$ is the angle between $\vec{r}$ and $\vec{B}$. The second distribution harmonic is given by
\begin{eqnarray}
& &   \int_0^{2\pi} f_{in}(\phi_p)\cos(2\phi_p)d\phi_p = 
    \frac{-2\pi}{p_T^2}\int bdbrdrd\phi_bd\phi_r\, {\cal{I}}_2, \nonumber \\ 
& & {\cal{I}}_2 = J_2(p_Tr)\cos(2\phi_r)  
    \nabla_r^2 S_A(r,b)\nabla_r^2 S_A(r,b'),
\end{eqnarray}
where $\phi_b$ is the angle between $\vec b$ and $\vec B$. In addition, the integral over $\phi_p$ is expressed as,
\begin{eqnarray}
&&   \int_0^{2\pi} f_{in}(\phi_p)d\phi_p =   
    \frac{2\pi}{p_T^2}\int bdbrdrd\phi_bd\phi_r\,{\cal{I}}_0 ,\nonumber \\ 
&& {\cal{I}}_0 = J_0(p_Tr) \nabla_r^2 S_A(x_1,r,b)\nabla_r^2 S_A(x_2,r,b').
\end{eqnarray}
The gluon jet decay with mass $m_j$ models the hadron production with transverse momentum $p_{T_h}$, where the hadron carries the average momentum fraction $\left<z\right>$. The initial produced hadron spectrum can be written as
\begin{eqnarray}
\label{eq:hadron}
f_{in}(p_{T_h},\phi_p)=\frac{K}{\left< z\right>^2}f_{in}\left(p_T^2=p_{T_h}^2/\left<z\right>^2+m_j^2,\phi_p\right).
\end{eqnarray}
Here the parameters $K$, $\left<z\right>$, and $m_j$ were determined for pion production in $pp$ collisions by Moriggi, Peccini, and Machado for different energy collisions \cite{Moriggi:2020zbv}. 

Using the Boltzmann-Gibbs-Blast-Wave (BGBW) model for produced hadrons in thermal equilibrium, we could model the collective flow and subsequent hydrodynamic expansion \cite{PhysRevC.48.2462}. This phenomenological model takes into account the main characteristics of hydrodynamic evolution. Thus we consider a velocity profile $\rho_0=tanh^{-1}(\beta_r)$ (with $\beta_r = \beta_s(\xi)^n$), determined by the surface expansion velocity $\beta_s$. Here, $\beta_r$ is the radial flow with $\xi = r/R_0$ being the ratio between the radial distance of the transverse plane and the fireball radius $R_0$. It is assumed a linear velocity profile, i.e. $n=1$. The azimuthal asymmetry can be parameterized by modulation in the velocity profile $\rho=\rho_0+\rho_a \cos(2\phi_s)$, where $\phi_s$ is the azimuthal angle in relation to reaction plane, as proposed by Huovinen et al \cite{Huovinen:2001cy}. The quantity $\rho_a$ is the anisotropy parameter in the flow. We observe the need to incorporate an extra parameter $s_2$ to describe the small $p_{T_h}$ region, whose purpose is introduce an azimuthal variation of the source density as described by STAR Collaboration at RHIC \cite{STAR:2001ksn}. Hence, the equilibrium distribution will be given by

\begin{eqnarray}\label{eq:BGBW}
    f_{eq}(\phi_p)\propto m_{T_h}\int_0^{R_0} rdr\int_0^{2\pi} d\phi_s K_1\left(\frac{m_{T_h}}{T_{eq}}\cosh(\rho(\phi_s))\right)  \nonumber \\
   \exp\left(\frac{p_{T_h}}{T_{eq}}\sinh(\rho(\phi_s))\cos(\phi_s-\phi_p)\right)\left[ 1+2s_2\cos(2\phi_s)\right], \nonumber \\
\end{eqnarray}
where $m_{T_h}=\sqrt{p_{T_h}^2+m_h^2}$ is the transverse mass of produced hadron  and $K_1$ is the modified Bessel function of second kind. The model above assumes that elliptic flow from collective origin is generated by a blending  of an azimuthal velocity
alteration and a spatially anisotropic freeze-out hyper-surface. In the expression above Bjorken correlation in rapidity, $\eta = y$, is assumed where $\eta $ is the space-time rapidity.

The equilibrium temperature $T_{eq}$ (in units of GeV) and the dimensionless parameters $\beta_s$, $\rho_a$, $s_2$ are  dependent on the collision impact parameter, $\vec{B}$, since the final momentum asymmetry must be dependent on the initial geometry  of the nuclear overlapping area. Such quantities are fitted from experimental data for each centrality class regarding $v_2$ and $\frac{dN_{AA}}{dp_{T_h}dy}$. As already mentioned, the final $p_T$ spectrum can be obtained from RTA-BTE approach \cite{Tripathy:2017kwb},
\begin{equation}
f_{fin}(\phi_p)=f_{eq}(\phi_p)+\left[f_{in}(\phi_p)-f_{eq}(\phi_p)\right]e^{-t_f/t_r},  
\end{equation}
where $t_r$ and $t_f$ are respectively the relaxation and freeze-out time. The initial hard distribution $f_{in}$, given by the Eqs. \eqref{eq:fatkt} and \eqref{eq:hadron}, evolves until reach the equilibrium distribution given by Eq. \eqref{eq:BGBW}, where the ratio $t_f/t_r$ is fitted for each centrality class. The second harmonic of $f_{fin}(\phi_p)$ gives the final elliptic flow coefficient $v_2$, which can be written as,
\begin{equation}\label{eq:v2}
    v_2=\frac{\int_0^{2\pi} f_{fin}(\phi_p)  \cos(2\phi_p)d\phi_p}{\int_0^{2\pi} f_{fin}(\phi_p)d\phi_p}.
\end{equation}

It is worth pointing out that not only the azimuthal asymmetry of the initial distribution has its origin in the dipole orientation, that depends on the impact parameter, but also generates a momentum asymmetry of Fourier Transform given by the integration shown in Eq. \eqref{eq:fatkt}. Meanwhile, in the equilibrium equation given by the Eq. \eqref{eq:BGBW}, the asymmetry can arise from the geometry of the initial collision, even with a vanishing second harmonic of the initial distribution.  An important point is that in our approach nuclear effects (nuclear shadowing) are already present in the nuclear UGD. It was shown in Refs. \cite{Moriggi:2020qla,Moriggi:2022xbg} that the nuclear shadowing effect changes significantly the spectrum at low-$p_T$ and consequently alters the fitted parameters in the  equilibrium distribution.

The eccentricity $\varepsilon_2$ of the initial collision can be obtained by the integration of Eq. \eqref{eq:hadron} with respect to transverse momentum $p_{T_h}$. We shall write this result as
\begin{equation}
\varepsilon_2 = \frac{\left< b_y^2-b_x^2\right>}{\left< b_y^2+b_x^2\right>},
\end{equation}
where $b_x$ is the component of $\vec{b}$ in the $\vec{B}$ direction.

In the present analysis, we consider only the spectrum of produced pions due the fact that its distribution is well described up to $p_{T_h}\lesssim 10$ GeV for LHC energies in $pp$ collisions \cite{Moriggi:2020zbv}. We need a good description of  the region  $p_{T_h}\gtrsim 5$ GeV to find appropriate values of the yield, Eq. \eqref{eq:NAA},  from a suitable parameter fitting. In this region the thermal distribution of equilibrium, given by Eq. \eqref{eq:BGBW}, is very small. The opposite scenario results in an unrealistic increase of $T_{eq}$ and $\beta_s$ values to fit the spectrum in the large $p_{T_h}$ region. After these consideration, we can write down the hadron productions in nuclear collision as

\begin{equation}
\label{eq:NAA}
 \frac{dN_{AA}}{p_{T_h}dp_{T_h}dy d\phi_p}=e^{-t_f/t_r}f_{in}(\phi_p)+(1-e^{-t_f/t_r})f_{eq}(\phi_p),
\end{equation}
and the nuclear modification factor will be given by,
\begin{equation}
\label{eq:RAA}
R_{AA}=\frac{\frac{d^3N_{AA}}{d\vec{p}^3}}{\langle T_{AA}\rangle \frac{d^3\sigma_{pp}}{d\vec{p}^3}},
\end{equation}
where $\langle T_{AA} \rangle $ is the average value of the nuclear overlapping function.

\begin{table}[]
\begin{tabular}{llllllll}
\hline
                          & centrality(\%) & $T_{eq}\,(\mathrm{GeV})$ & $\beta_s$ & $\rho_a$ & $s_2$  & $t_f/t_r$ & $\chi^2 /dof$ \\ \hline
\multicolumn{1}{l|}{PbPb} & 05-10           & 0.229    & 0.504     & 0.0386   & 0.0590 & 2.05       & 0.990         \\
\multicolumn{1}{l|}{}     & 10-20          & 0.148    & 0.787     & 0.0678   & 0.0670 & 1.86       & 1.126         \\
\multicolumn{1}{l|}{}     & 20-30          & 0.122    & 0.856     & 0.103    & 0.0835 & 1.64       & 1.389         \\
\multicolumn{1}{l|}{}     & 30-40          & 0.116    & 0.867     & 0.128    & 0.110  & 1.38       & 1.488         \\
\multicolumn{1}{l|}{}     & 40-50          & 0.100    & 0.900     & 0.167    & 0.132  & 1.17       & 1.683         \\
\multicolumn{1}{l|}{}     & 50-60          & 0.090    & 0.911     & 0.202    & 0.164  & 0.89       & 1.122         \\
\multicolumn{1}{l|}{}     & 60-70          & 0.091    & 0.910     & 0.210    & 0.254  & 0.36       & 0.766         \\ \hline
\multicolumn{1}{l|}{XeXe} & 05-10           & 0.190    & 0.671     & 0.0351   & 0.057  & 2.03       & 0.556         \\
\multicolumn{1}{l|}{}     & 10-20          & 0.121    & 0.862     & 0.0782   & 0.065  & 1.65       & 0.232         \\
\multicolumn{1}{l|}{}     & 20-30          & 0.131    & 0.829     & 0.0941   & 0.106  & 1.29       & 0.167         \\
\multicolumn{1}{l|}{}     & 30-40          & 0.116    & 0.868     & 0.129    & 0.129  & 1.02       & 0.105         \\
\multicolumn{1}{l|}{}     & 40-50          & 0.125    & 0.843     & 0.113    & 0.187  & 0.68       & 0.250         \\
\multicolumn{1}{l|}{}     & 50-60          & 0.151    & 0.767     & 0.0771   & 0.324  & 0.31       & 0.098        
\end{tabular}
\caption{Kinetic freeze-out parameters in each centrality class for production of charged pions in PbPb and XeXe collisions at $\sqrt{s}=5.02$ TeV and $\sqrt{s}=5.44$ TeV at the LHC, respectively. Parameters $\beta_s,\,\rho_a,\,s_2$ and $t_f/t_r$ are dimensionless.  Only central values are shown and the errors are typically of 20\%.} 
\label{tab:pars}
\end{table}

\section{Results and discussion}
\label{sec:results}

\begin{table*}[t]
\begin{tabular}{llllllll}
      Centrality                             & 5-10             & 10-20             & 20-30              & 30-40            & 40-50            & 50-60             & 60-70            \\ \hline
 \multicolumn{1}{l|}{$PbPb$} & $0.668\pm 0.068$ & $0.261\pm 0.015$  & $0.151\pm0.086 $   & $0.130\pm 0.075$ & $0.113\pm 0.073$ & $0.114\pm 0.01$   & $0.114\pm 0.015$ \\ 
       \multicolumn{1}{l|}{$XeXe$} & $0.70 \pm 0.42$  & $0.190 \pm 0.054$ & $0.1809 \pm 0.033$ & $0.128\pm 0.031$ & $0.171\pm 0.038$ & $0.189 \pm 0.057$ &                 
\end{tabular}
\caption{Results of fitting the dimensionless parameter  $a_r$ at different centralities for PbPb and XeXe collisions.}
\label{tab:ar}
\end{table*}

\begin{figure*}[t]
\includegraphics[width=0.7\linewidth]{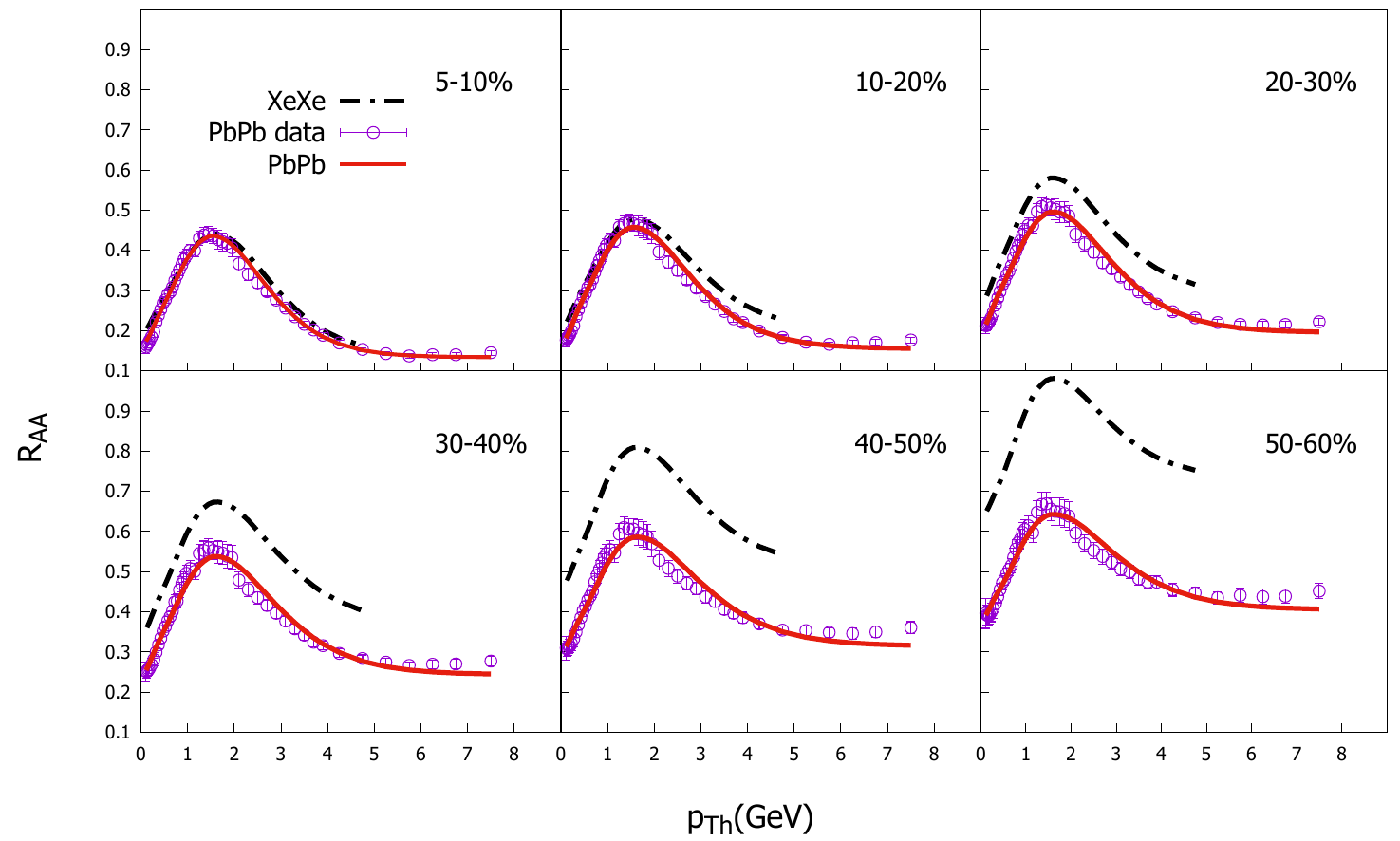}
\caption{Nuclear modification factors for PbPb collisions compared with data of ALICE collaboration \cite{ALICE:2019hno} for different centrality classes. The results for XeXe collisions are are also presented.} 
\label{fig:RAA} 
\end{figure*}

We now proceed to the analysis of the available data at the LHC for the nuclear modification factor and elliptic flow by using the theoretical approach presented in previous section. The  $p_{T_h}$ spectrum and $v_2$ coefficient were fitted in the interval $0.1<p_{T_h}<8$ GeV for data of ALICE collaboration \cite{ALICE:2018yph,ALICE:2019hno,ALICE:2021ibz,ALICE:2021lsv}, regarding pion production in PbPb and XeXe collisions at $\sqrt{s}=5.02$ TeV and $\sqrt{s}=5.44$ TeV, respectively. Since that initial distribution given by Eq. \eqref{eq:fatkt} is not capable of generating enough asymmetry to describe $v_2$ in these very central collisions, we excluded the centrality class of 0-5\% in our analysis. The BGBW distribution parameters and the ratio $t_f/t_r$ are shown in the Table \ref{tab:pars}. The Woods-Saxon nuclear density was considered to compute the nuclear thickness function $T_A(b)$, using parameters from De Vries et al \cite{DEVRIES1987495}. It is worth mentioning that in the present analysis we did not consider the deformation of Xe nuclei.

Thus, this set of parameters are in good agreement with the expected behavior where $T_{eq}$ decreases with the centrality, whereas the quantities that define azimuthal asymmetry $\rho_a$ and $s_2$ increase with the centrality.  In order to discuss the freeze-out parameters, for PbPb collisions it is shown that $\langle \beta_r \rangle = \left( \frac{2}{2+n}\right)\beta_s $ ($\langle \beta_r \rangle = (2/3)\beta_s$, for $n=1$) decreases with centrality, reaching  $\langle \beta_r \rangle =  0.336\pm 0.024$ in $5-10\%$ central
collisions, while the equilibrium temperature increases going from  $T_{eq} = (0.091 \pm 0.015)$ GeV to $T_{eq} = (0.229 \pm 0.084)$ GeV.   As considered in a previous analysis 
 \cite{Moriggi:2020qla}, the radial velocity $\beta_s$ is anti-correlated with $T_{eq}$ and therefore diminishes with the centrality. Although the values of $\chi^2/dof$ shows sizable deviation from the unit for the PbPb case, the resulting parameters have values physically consistent. Namely, the transverse momentum spectra data cover greater values of $p_{T_h}$ and they help to restrict $T_{eq}$ and $\beta_s$, generating smaller errors in the evaluation of these parameters. It should be stressed that the value of the exponent of the expansion velocity profile, $n$, is equal unity in the present case. The situation is different in the fits of spectra with a blast-wave function done by ALICE Collaboration in Refs. \cite{ALICE:2013mez,ALICE:2019hno} where $n$ is about 0.74 in central collisions and it increases up to 2.52 in more peripheral collisions. Here, a variation of $n$ was not necessary to reproduce the large-$p_T$ tail of the spectra as it is not thermal over the full range of transverse momentum.

 In general, the centrality dependence of the thermal parameters $T_{eq}$ and $\beta_s $ appearing in Tab. \ref{tab:pars} has the opposite behavior compared to the BGBW approaches \cite{Tripathy:2017nmo,Younus:2018mrk,Huovinen:2001cy}. The main reason is that the shadowing/anti-shadowing effect in the small-$p_T$ region alters the corresponding particle distribution and modifies the parameter associated to equilibrium temperature. Let us take for instance a Boltzmann distribution $\sim  \exp(-p_T/T)$: the temperature quantifies how much disperse the $p_T$ spectrum is. On the other hand, the initial state nuclear effects also produce a broadening of that distribution. In more peripheral collisions the hard (initial) distribution starts to become more prominent and its effect is almost sufficient to describe particle spectrum. In this sense, the temperature can be smaller in this case compared to more central collisions.

The dimensionless parameter $a_r$ associated to azimuthal modulation in the dipole cross section is more relevant in the large $p_T$ region. In the  $p_{T_h} \sim 5$ GeV region the distribution $f_{eq}$ is very small and the final asymmetry is basically related to $f_{in}$ and therefore the parameter $a_r$ can be determined separately. Further, we defined $v_{2in}$ associated only with $f_{in}$, which was fitted within the region of $p_{T_h} > 5$ GeV. The results are shown in the Table \ref{tab:ar}. We can see that in more central collisions (5-10\%) the  asymmetry generated by the initial distribution is very small and $a_r$ should be large to fit the data. For bigger values of centrality, $a_r$ is almost constant, with $a_r\approx 0.18$. At this stage of the phenomenological study it is not clear how to explain the dependence of $a_r$ on the centrality. Based on the work of Ref. \cite{Iancu:2017fzn}, the nuclear color dipole amplitude with dipole orientation can be written as $N_{2g}(r,B,\phi_{rb}) = {\cal{N}}_0^A(r,B)\left[ 1+\kappa (B)\cos(2\phi_{rb} \right]$, with $\kappa (B)\sim [T_A^{\prime\prime}(B)-T_A^{\prime}(B)/B]/T_A(B)$. Here, $T_A^{\prime \prime}$ and $T_A^{\prime}$ are the second and first derivative of the thickness function, respectively. Using by simplification of a Gaussian thickness $T_A\propto \exp (-B^2/R_A^2)$, with $R_A$ being the nuclear radius, one obtains $\kappa (B)\sim (2B/R_A^2)^2$ meaning that the asymmetry is more intense for peripheral collisions.  This feature is also confirmed by the numerical solution of BK evolution equation when dipole orientation is included \cite{Golec-Biernat:2003naj,PhysRevD.83.034015}.

\begin{figure*}[t]
\includegraphics[width=0.7\linewidth]{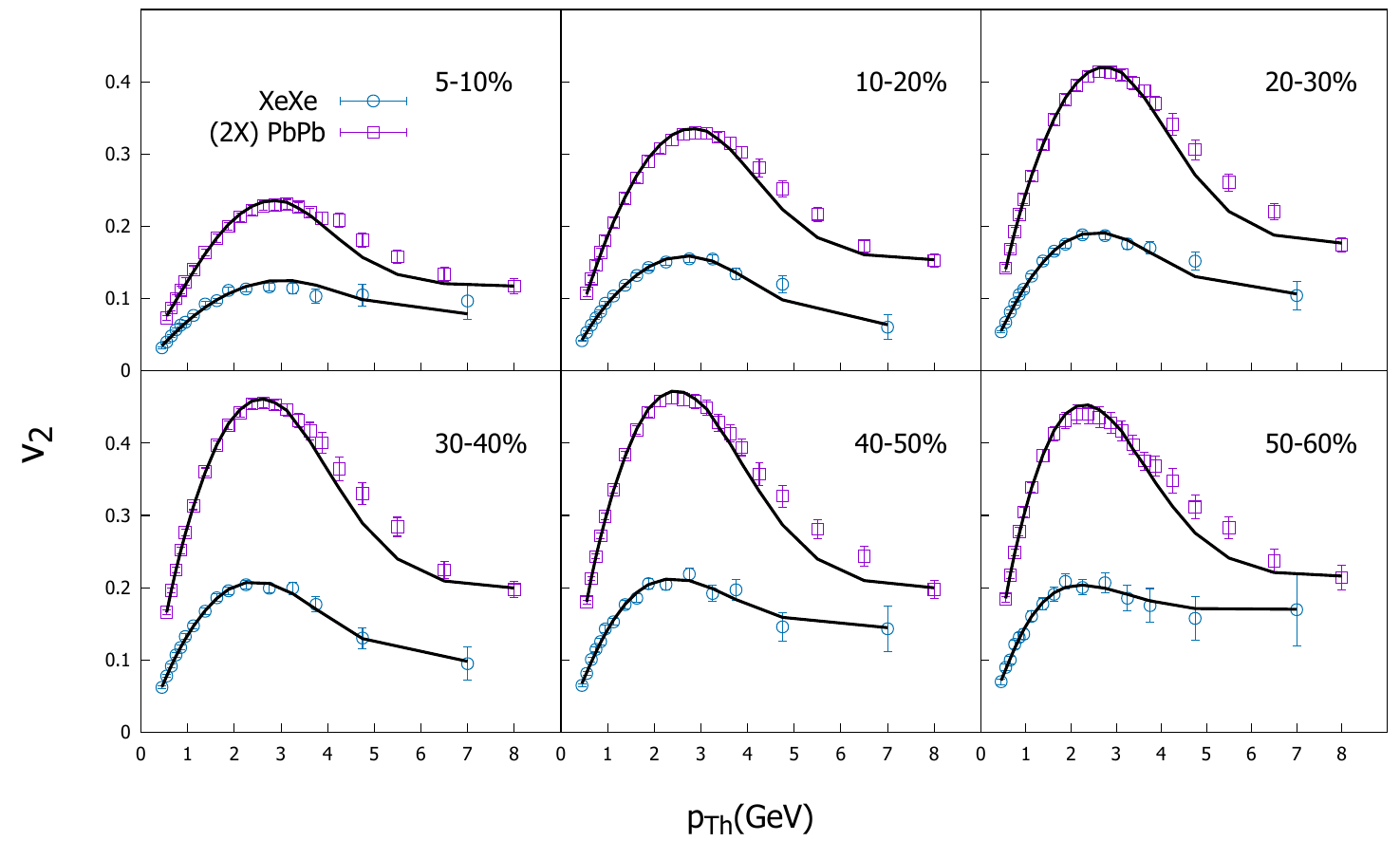}
\caption{Elliptic flow coefficient $v_2(p_{T_h})$ for XeXe and PbPb collisions compared to  data from ALICE collaboration \cite{ALICE:2021ibz,ALICE:2018yph}. The results for PbPb have been multiplied by factor 2 for a better visualization.}
\label{fig:v2} 
\end{figure*}

We present in the Figure \ref{fig:NAA} the final spectrum, given by the Eq. \eqref{eq:NAA}, in comparison with experimental data of ALICE collaboration for XeXe and PbPb collisions. The spectrum of produced pions in $pp$ collisions is also shown. It is important to note that we need a good agreement with the $pp$ comparison case to have a suitable description of nuclear modification factor $R_{AA}$. If this is not the case, we can still have a good fit of multiplicity but not a suitable description of $R_{AA}$. In the Figure \ref{fig:RAA}, the resulting nuclear modification factor from the fitting is compared with data from ALICE collaboration \cite{ALICE:2019hno} of charged pion production. The dashed lines are the fitting results of the spectrum for XeXe collision evaluated with the Eq. \eqref{eq:RAA}. For central collisions, $R_{AA}$ is basically equal for PbPb and XeXe. In our model, this is due mainly to the fact that in both cases the ratio $t_f/t_r$ is essentially the same, implying that both systems evolve in a similar time. In more peripheral collisions, the ratio $t_f/t_r$ for XeXe is remarkably smaller, which generates less suppression of initial spectrum.

\begin{figure*}[t]
\includegraphics[width=0.7\linewidth]{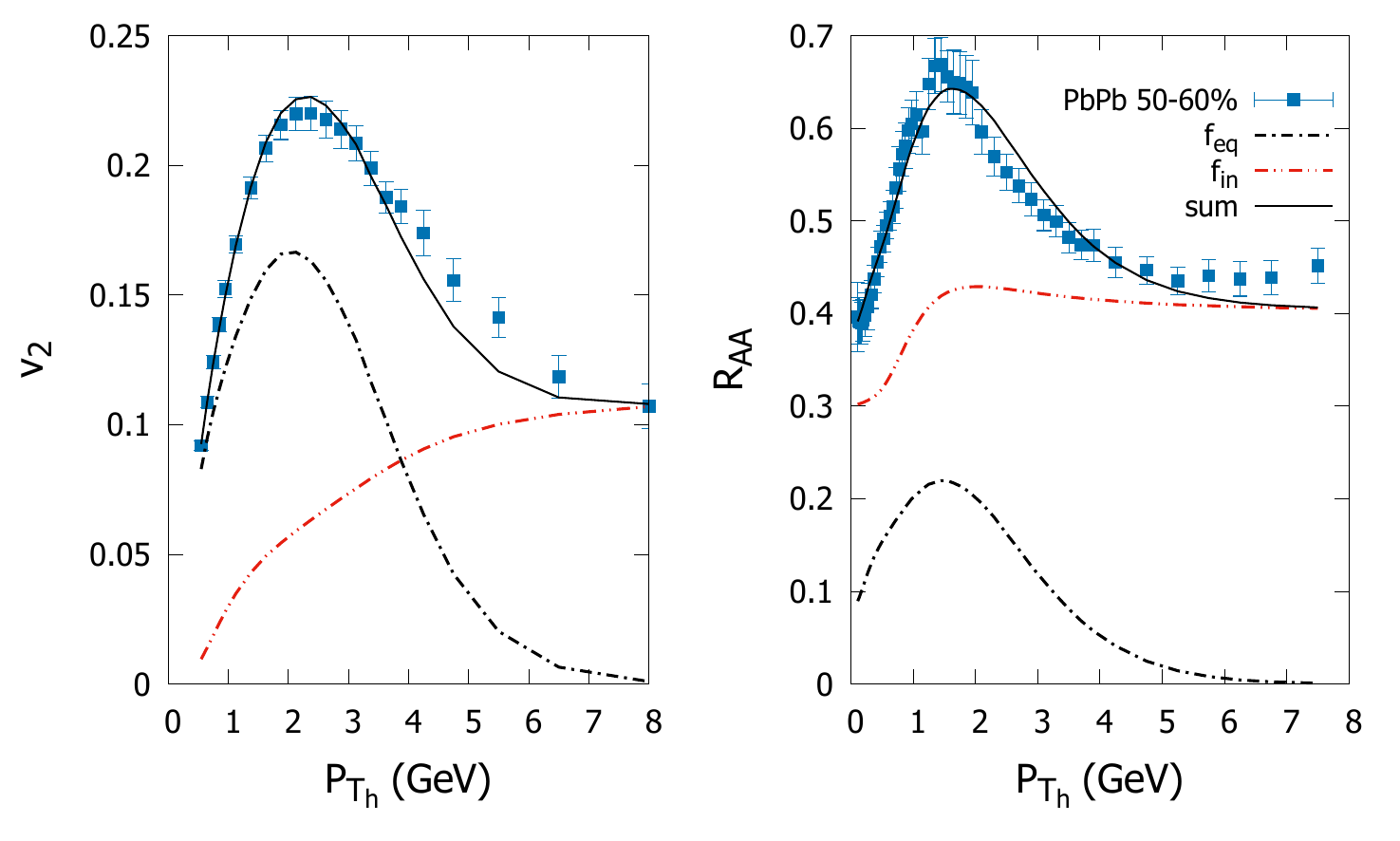}
\caption{Detailed analysis for the $f_{in}$ and $f_{eq}$ contributions for the final asymmetry $v_2$ (left plot) and nuclear modification $R_{AA}$ (right plot) for PbPb collision at centrality (50-60)\%.}
\label{fig:finfeq} 
\end{figure*}

The differential elliptic flow $v_2(p_{T_h})$ evaluated by Eq. \eqref{eq:v2} is shown in the Figure \ref{fig:v2}. The results are contrasted with data for XeXe and PbPb collisions in several centralities. We can see that our model has better agreement with $v_2$ for XeXe case, where $\chi^2$ is smaller than the PbPb one. At a given 
 value of $p_{T_h}$, the more peripheral collisions have the largest value of elliptic flow, and $v_2$ decreases for more central collisions. For all the centralities, in the small $p_{T_h}$ range, the transverse momentum dependence of the flow for charged pions is approximately linear.

The interface between the initial and final distribution is illustrated by Figure \ref{fig:finfeq}, where the nuclear modification factor $R_{AA}$ (plot on the right) and $v_2$ (plot on the left) as function of $p_{T_h}$ are shown. In these results, the dashed lines represent the $f_{in}$ (curves in color red) and $f_{eq}$ (curve in color black) contributions and the solid lines the summation. To be clear,  Fig. \ref{fig:finfeq} shows the relative contribution to the $p_T$-spectrum coming from initial and equilibrium  distributions appearing in Eq. (\ref{eq:NAA}). Namely, it is shown the contributions driven by $f_{in}(p_T)\times e^{-t_f/t_r}$ and  $f_{eq}\times (1-e^{t_f/t_r})$ terms. The nuclear modification factor $R_{AA}$ is given by the sum of these two contributions. Both $R_{AA}$ and $v_2$ grows up to their highest values for $p_{T_h} \sim 2$ GeV, where the $f_{eq}$ contribution is maximum. After this region, $R_{AA}$ and $v_2$ decreases in value, indicating a predominance of the contribution of produced particles at initial stages of the collision. In our model, the growth of $v_2$ and the Cronin peak has a common origin as the interface between these two distributions. It also indicates a limit to our approach, since that for $p_{T_h} \gtrsim 10$ GeV $v_2$ decreases, while $R_{AA}$ increases. In this picture, models that consider energy loss or color transparency can be considered. Please, see \cite{Kopeliovich:2012sc,Kopeliovich:2021dgx,Zhao:2021vmu,Nemchik:2014gka} for further discussion on this topic. We can see that the azimuthal asymmetry of the momentum generated by initial collision is basically given by partons with large $p_T$, whereas the asymmetry in the equilibrium occurs mainly in small $p_T$.

\begin{figure}[t]
\includegraphics[width=\linewidth]{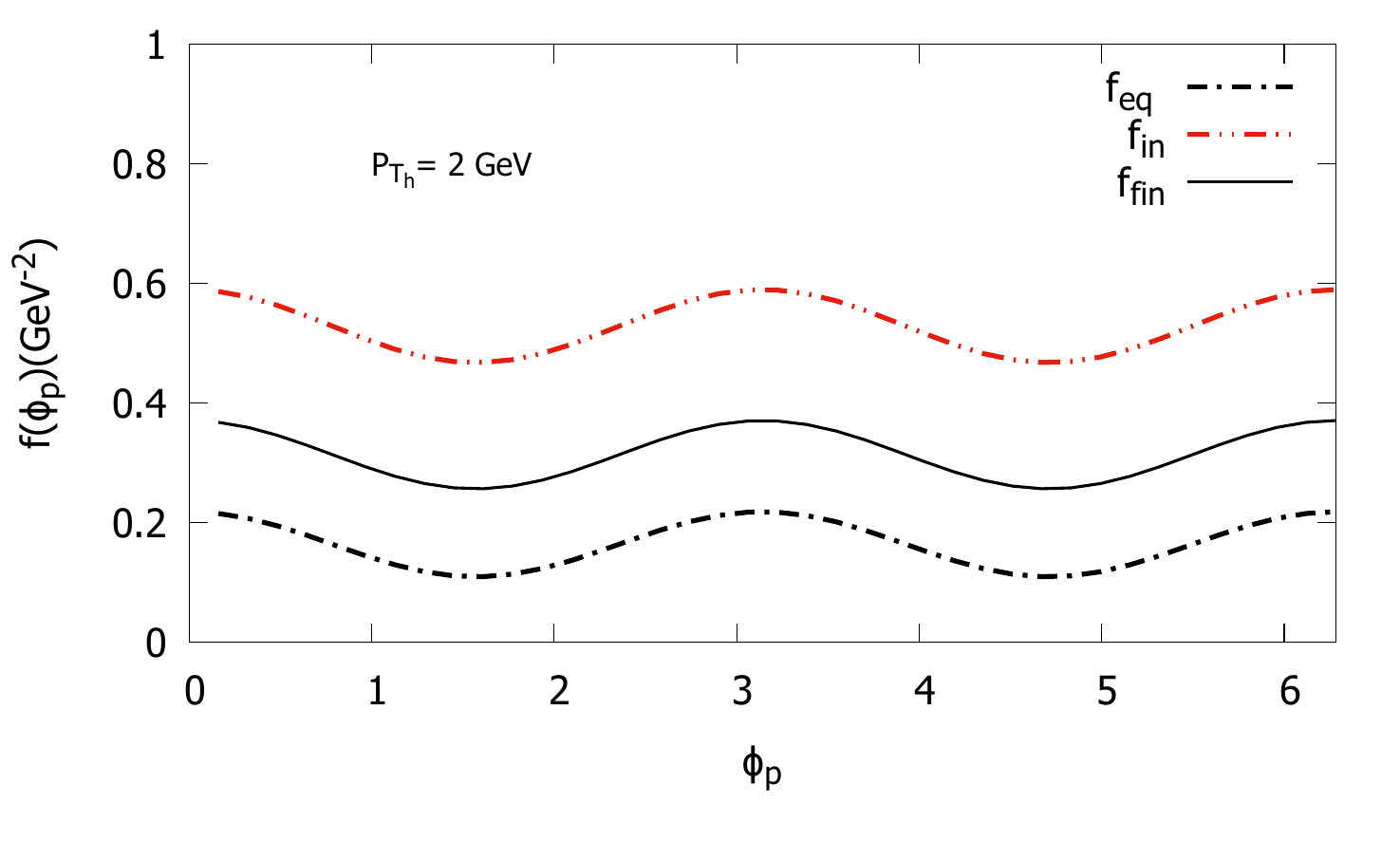}
\caption{The $\phi_p$ dependence of the initial, final and equilibrium distributions. Contributions for $p_T=2$ GeV and 50-60$\%$ centrality class in PbPb collisions are presented. Distributions $f_{in}$ and $f_{eq}$ are respectively labeled by the upper and lower dot-dashed curves  whereas $f_{fin}$ is given by the solid curve.}
\label{fig:finfeq-azimuth} 
\end{figure}

In order to illustrate the contributions arising from $f_{in}$ and $f_{eq}$ as a function of the angle between $\vec{p}_T$ and impact parameter $\vec{B}$, $\phi_p$, in Fig. \ref{fig:finfeq-azimuth} the corresponding distributions are shown. Accordingly, as in Fig. \ref{fig:finfeq} the distributions   $f_{in}$ (upper dot-dashed curve), $f_{eq}$ (lower dot-dashed curve) and $f_{fin}$ (solid line curve) are presented as a function of $\phi_p$ for  PbPb collisions at 50-60$\%$ centrality class. Transverse momentum $p_T=2$ GeV has been considered in which $f_{in}$ and $f_{eq}$ give similar order of magnitude contributions to the final distribution. The values of the final distribution, $f_{fin}$, are in the middle of the equilibrium $f_{eq}$ and initial $f_{in}$ cases.

In the remainder we investigated  the  behaviour of $v_2$ as a function of $p_{T_h}$ for different centralities and its
dependence on the anisotropies generated at initial states and from the generalization of the blast wave model. Before going into
the conclusions  we discuss additional observables sensitive to the ingredients considered in our approach. As examples, the ratio $v_2/\varepsilon_2$ and the average transverse momentum of particles are discussed.

\begin{figure}[t]
\includegraphics[width=\linewidth]{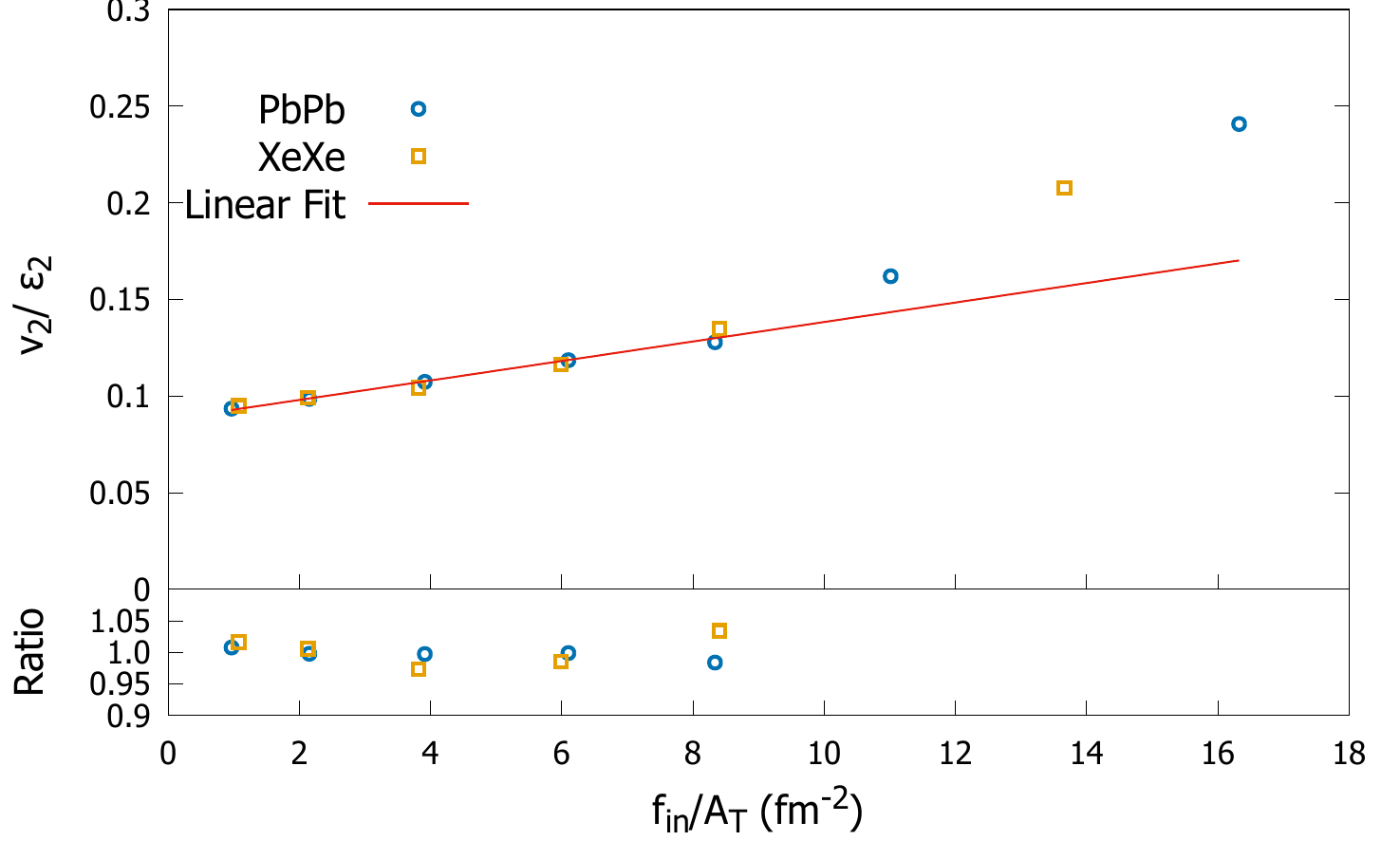}
\caption{Ratio $v_2/\varepsilon_2$ as function of $f_{in}/A_T$ for PbPb and XeXe at different centralities. The solid line represents the linear fit following Eq. \ref{eq:scaling1}. Plot in the bottom shows the ratio data/model.}
\label{fig:v2_sc} 
\end{figure}

In the work of Voloshin and Poskanze \cite{Voloshin:1999gs}, is proposed that the ratio of final elliptic flow with the initial anisotropy, $v_2/\varepsilon_2$, can indicate the equilibrium level of produced system. In the limit of low density one expects
\begin{equation}\label{eq:scaling1}
    \frac{v_2}{\varepsilon_2} \propto \frac{1}{A_T}f_{in},
\end{equation}
where $A_T$ is the transverse area given by the region of nuclear overlapping area and $f_{in}$ is the integrated initial distribution in $p_T$. On the other hand, in the hydrodynamic limit, representing the complete thermalization of the final system, we should have $v_2 \propto \varepsilon_2$. Therefore, scaling deviation given by the Eq. \eqref{eq:scaling1} can indicate different mechanisms of particle production. The centrality dependency of $v_2$ is considered for each centrality by integrating $v_2$ in $p_{T_h}$,

\begin{equation}
    v_2(B)=\frac{\int dp_{T_h} p_{T_h} v_2(p_{T_h}) f_{fin}}{\int dp_{T_h} p_{T_h} f_{fin}}.
\end{equation}

The elliptic flow replicates the space and momentum correlation developed because of early stage pressure gradient. On the other hand, $A_T^{-1} f_{in}$ can be associated to a measure of the transverse particle density. Therefore, the plot $v_2/\varepsilon_2\times A_T^{-1}f_{in}$ could be viewed as an analogous of pressure versus energy density plot. The Figure \ref{fig:v2_sc} the variation of eccentricity scaled elliptic flow with such transverse density. It shows the ratio $v_2/\varepsilon_2$ as function of $f_{in}/A_T$, with $f_{in}$ integrated over $p_{T_h}$ for different centralities for both PbPb (open circles) and XeXe collisions (open squares). The expected linear scaling (solid line in figure) is a good approximation in more peripheral collisions. For large multiplicities, there is a substantial deviation of the fitted line. In our model, such effect results of the increasing in $f_{eq}$ contribution for the spectrum in more central collisions.

\begin{figure}[t]
\includegraphics[width=\linewidth]{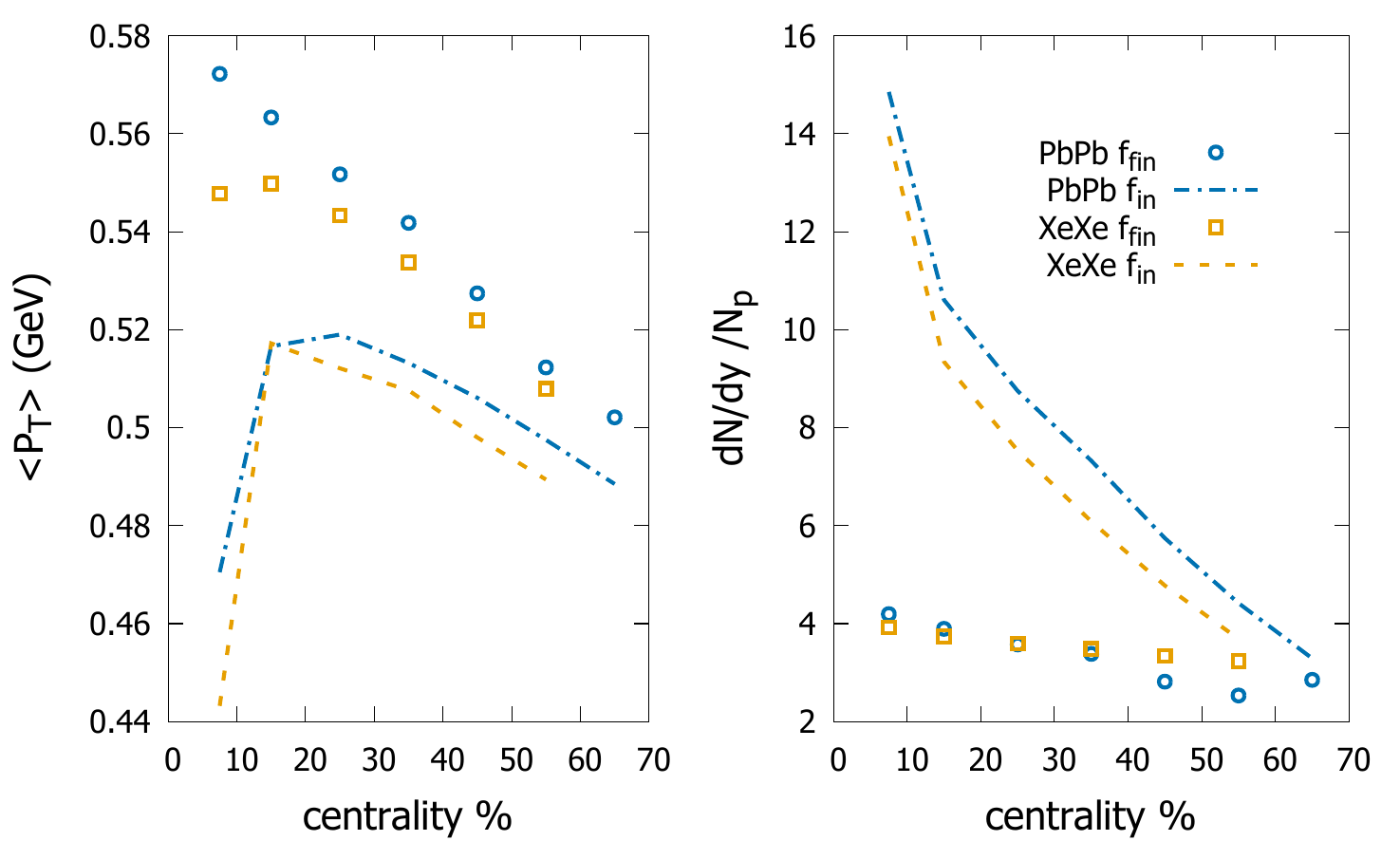}
\caption{(Left) Initial and final mean transverse momentum $\langle p_{T_h}\rangle$ for different centralities. (Right) Initial and final ratio $\frac{dN_{AA}/dy}{N_p}$ (right) as a function of centrality.} 
\label{fig:ptm} 
\end{figure}

Finally, the initial and final mean transverse momentum is presented in Figure \ref{fig:ptm} (left plot) for different centralities. Initial $\langle p_{T_h}\rangle$ are represented by dot-dashed line for PbPb and dashed line for XeXe, whereas final $\left< p_{T_h}\right>$ is represented by open circles for PbPb and open squares for XeXe. For more central collisions, there is a large increasing of $\left< p_T\right>$ due the temperature increasing of the distribution $f_{eq}$. However, for more peripheral collisions, the final transverse momentum comes closer of initial transverse momentum. The multiplicity by wounded nucleon is also shown. While the final distribution scales with $N_p$, $f_{in}$ grows faster than $N_p$ in more central collisions.

We did not discuss universality aspects related to the elliptic flow in the context of parton saturation physics used in our theoretical formalism for $f_{in}$. Interestingly, it was demonstrated in Refs. \cite{Andres:2014xka,AndresCasas:2017glq}  that the available data on the $v_2$ of charged particles for collisions at RHIC and LHC for different centralities present a scaling law. Namely, the elliptic flow normalized by the product of the inverse of the Knudsen number, $\mathrm{Kn}^{-1}=Q_{s,A}L$, and eccentricity $\epsilon_1 $ satisfy geometrical scaling. Here, $Q_{s,A}=Q_{s,A}(x,B;A) $ is the nuclear saturation scale and $L$ is the length related to the size of the collision area. The normalized $v_2$ is a function only the $\tau_A = p_{T_h}^2/Q_{s,A}^2$ variable,
\begin{eqnarray}
\frac{v_2(p_{T_h})}{\epsilon_1/\mathrm{Kn}} = F(\tau_A)=a\tau_A^b,
\end{eqnarray}
with the parameters $a$ and $b\sim 1/3$ fitted from data \cite{Andres:2014xka,AndresCasas:2017glq} and $0\leq \tau_A\leq 1$. It is a relative hard task to derive such a scaling law  from the geometrical scaling phenomenon. It is argued in \cite{Andres:2016mla} that the scaling law can be traced back to the parton energy lost in nucleus-nucleus collisions.

\section{Summary and conclusions}
\label{sec:conc}

Let us now organize the main results and draw some
conclusions. In this work we presented a phenomenological model that incorporates the main characteristics of produced particle spectrum in heavy ions collisions. For evaluate the inclusive gluon production, it has been considered hard initial collision, nuclear shadowing of partonic distribution, and azimuthal asymmetry in this model. Also the process makes use of dipole scattering matrix in the position space, where an initial asymmetry has been added. However, collective effects are described by BGBW distribution, leading to an increasing of the momentum asymmetry for small $p_{T_h}$. One advantage of the present parameterization is that it allow us to describe an interface between different mechanisms of hadron production providing a good description of both soft and hard contributions. In particular, we argue that the behavior of $v_2$ and $R_{AA}$ has a common origin, i.e., the increasing in the contribution of produced particles in thermal equilibrium and the subsequent predominance of the distribution  of produced particle at initial hard peripheral collisions. The nuclear modification factor for pion production in PbPb and XeXe collisions and respective elliptic flow $v_2(p_T)$ have been simultaneously described. The fitted parameters $T_{eq}$, $\beta_s$, $\rho_a$ and $s_2$ are associated to the modified BGBW approach whereas the parameter $a_r$ is connected to the azimuthal asymmetry in the early stages of the collision. The quality of fit is good and the role played by $f_{in}$ and $f_{eq}$ in description of $R_{AA}$ and $v_2$ has been discussed. Finally, the ratio $v_2/\varepsilon_2$ as function of $f{in}/A_T$ is computed and analyzed in terms of linear scaling between these two quantities. The average transverse momentum, $\langle p_T\rangle$, has been also investigated.

\section*{Acknowledgments}
This research was funded by the Brazilian National Council for Scientific and Technological Development (CNPq) under the contract number 303075/2022-8.

\bibliographystyle{apsrev4-2}
\bibliography{referenciasAA}

\begin{thebibliography}{82}%
\makeatletter
\providecommand \@ifxundefined [1]{%
 \@ifx{#1\undefined}
}%
\providecommand \@ifnum [1]{%
 \ifnum #1\expandafter \@firstoftwo
 \else \expandafter \@secondoftwo
 \fi
}%
\providecommand \@ifx [1]{%
 \ifx #1\expandafter \@firstoftwo
 \else \expandafter \@secondoftwo
 \fi
}%
\providecommand \natexlab [1]{#1}%
\providecommand \enquote  [1]{``#1''}%
\providecommand \bibnamefont  [1]{#1}%
\providecommand \bibfnamefont [1]{#1}%
\providecommand \citenamefont [1]{#1}%
\providecommand \href@noop [0]{\@secondoftwo}%
\providecommand \href [0]{\begingroup \@sanitize@url \@href}%
\providecommand \@href[1]{\@@startlink{#1}\@@href}%
\providecommand \@@href[1]{\endgroup#1\@@endlink}%
\providecommand \@sanitize@url [0]{\catcode `\\12\catcode `\$12\catcode
  `\&12\catcode `\#12\catcode `\^12\catcode `\_12\catcode `\%12\relax}%
\providecommand \@@startlink[1]{}%
\providecommand \@@endlink[0]{}%
\providecommand \url  [0]{\begingroup\@sanitize@url \@url }%
\providecommand \@url [1]{\endgroup\@href {#1}{\urlprefix }}%
\providecommand \urlprefix  [0]{URL }%
\providecommand \Eprint [0]{\href }%
\providecommand \doibase [0]{http://dx.doi.org/}%
\providecommand \selectlanguage [0]{\@gobble}%
\providecommand \bibinfo  [0]{\@secondoftwo}%
\providecommand \bibfield  [0]{\@secondoftwo}%
\providecommand \translation [1]{[#1]}%
\providecommand \BibitemOpen [0]{}%
\providecommand \bibitemStop [0]{}%
\providecommand \bibitemNoStop [0]{.\EOS\space}%
\providecommand \EOS [0]{\spacefactor3000\relax}%
\providecommand \BibitemShut  [1]{\csname bibitem#1\endcsname}%
\let\auto@bib@innerbib\@empty
\bibitem [{\citenamefont {{Arslandok}}\ \emph {et~al.}(2023)\citenamefont
  {{Arslandok}} \emph {et~al.}}]{Arslandok:2023utm}%
  \BibitemOpen
  \bibfield  {author} {\bibinfo {author} {\bibfnamefont {M.}~\bibnamefont
  {{Arslandok}}} \emph {et~al.},\ }\href {\doibase 10.48550/arXiv.2303.17254}
  {\bibfield  {journal} {\bibinfo  {journal} {arXiv e-prints}\ ,\ \bibinfo
  {eid} {arXiv:2303.17254}} (\bibinfo {year} {2023})},\ \Eprint
  {http://arxiv.org/abs/2303.17254}{arXiv:2303.17254 [nucl-ex]}\BibitemShut
  {NoStop}%
\bibitem [{\citenamefont {Shuryak}(1978)}]{Shuryak:1978ij}%
  \BibitemOpen
  \bibfield  {author} {\bibinfo {author} {\bibfnamefont {E.~V.}\ \bibnamefont
  {Shuryak}},\ }\href {\doibase 10.1016/0370-2693(78)90370-2} {\bibfield
  {journal} {\bibinfo  {journal} {Phys. Lett. B}\ }\textbf {\bibinfo {volume}
  {78}},\ \bibinfo {pages} {150} (\bibinfo {year} {1978})}\BibitemShut
  {NoStop}%
\bibitem [{\citenamefont {Shuryak}(1980)}]{Shuryak:1980tp}%
  \BibitemOpen
  \bibfield  {author} {\bibinfo {author} {\bibfnamefont {E.~V.}\ \bibnamefont
  {Shuryak}},\ }\href {\doibase 10.1016/0370-1573(80)90105-2} {\bibfield
  {journal} {\bibinfo  {journal} {Phys. Rept.}\ }\textbf {\bibinfo {volume}
  {61}},\ \bibinfo {pages} {71} (\bibinfo {year} {1980})}\BibitemShut {NoStop}%
\bibitem [{\citenamefont {Das}\ \emph {et~al.}(2022)\citenamefont {Das} \emph
  {et~al.}}]{Das:2022lqh}%
  \BibitemOpen
  \bibfield  {author} {\bibinfo {author} {\bibfnamefont {S.~K.}\ \bibnamefont
  {Das}} \emph {et~al.},\ }\href {\doibase 10.1142/S0218301322500975}
  {\bibfield  {journal} {\bibinfo  {journal} {Int. J. Mod. Phys. E}\ }\textbf
  {\bibinfo {volume} {31}},\ \bibinfo {pages} {12} (\bibinfo {year} {2022})},\
  \Eprint {http://arxiv.org/abs/2208.13440}{arXiv:2208.13440
  [nucl-th]}\BibitemShut {NoStop}%
\bibitem [{\citenamefont {Ritter}\ and\ \citenamefont
  {Stock}(2014)}]{Ritter:2014uca}%
  \BibitemOpen
  \bibfield  {author} {\bibinfo {author} {\bibfnamefont {H.~G.}\ \bibnamefont
  {Ritter}}\ and\ \bibinfo {author} {\bibfnamefont {R.}~\bibnamefont {Stock}},\
  }\href {\doibase 10.1088/0954-3899/41/12/124002} {\bibfield  {journal}
  {\bibinfo  {journal} {J. Phys. G}\ }\textbf {\bibinfo {volume} {41}},\
  \bibinfo {pages} {124002} (\bibinfo {year} {2014})},\ \Eprint
  {http://arxiv.org/abs/1408.4296}{arXiv:1408.4296 [nucl-ex]}\BibitemShut
  {NoStop}%
\bibitem [{\citenamefont {Ollitrault}(1992)}]{Ollitrault:1992bk}%
  \BibitemOpen
  \bibfield  {author} {\bibinfo {author} {\bibfnamefont {J.-Y.}\ \bibnamefont
  {Ollitrault}},\ }\href {\doibase 10.1103/PhysRevD.46.229} {\bibfield
  {journal} {\bibinfo  {journal} {Phys. Rev. D}\ }\textbf {\bibinfo {volume}
  {46}},\ \bibinfo {pages} {229} (\bibinfo {year} {1992})}\BibitemShut
  {NoStop}%
\bibitem [{\citenamefont {Voloshin}(2009)}]{Voloshin:2009fd}%
  \BibitemOpen
  \bibfield  {author} {\bibinfo {author} {\bibfnamefont {S.~A.}\ \bibnamefont
  {Voloshin}},\ }\href {\doibase 10.1016/j.nuclphysa.2009.05.082} {\bibfield
  {journal} {\bibinfo  {journal} {Nucl. Phys. A}\ }\textbf {\bibinfo {volume}
  {827}},\ \bibinfo {pages} {377C} (\bibinfo {year} {2009})}\BibitemShut
  {NoStop}%
\bibitem [{\citenamefont {Voloshin}\ and\ \citenamefont
  {Zhang}(1996)}]{Voloshin:1994mz}%
  \BibitemOpen
  \bibfield  {author} {\bibinfo {author} {\bibfnamefont {S.}~\bibnamefont
  {Voloshin}}\ and\ \bibinfo {author} {\bibfnamefont {Y.}~\bibnamefont
  {Zhang}},\ }\href {\doibase 10.1007/s002880050141} {\bibfield  {journal}
  {\bibinfo  {journal} {Z. Phys. C}\ }\textbf {\bibinfo {volume} {70}},\
  \bibinfo {pages} {665} (\bibinfo {year} {1996})}\BibitemShut {NoStop}%
\bibitem [{\citenamefont {Poskanzer}\ and\ \citenamefont
  {Voloshin}(1998)}]{Poskanzer:1998yz}%
  \BibitemOpen
  \bibfield  {author} {\bibinfo {author} {\bibfnamefont {A.~M.}\ \bibnamefont
  {Poskanzer}}\ and\ \bibinfo {author} {\bibfnamefont {S.~A.}\ \bibnamefont
  {Voloshin}},\ }\href {\doibase 10.1103/PhysRevC.58.1671} {\bibfield
  {journal} {\bibinfo  {journal} {Phys. Rev. C}\ }\textbf {\bibinfo {volume}
  {58}},\ \bibinfo {pages} {1671} (\bibinfo {year} {1998})}\BibitemShut
  {NoStop}%
\bibitem [{\citenamefont {Snellings}(2011)}]{Snellings:2011sz}%
  \BibitemOpen
  \bibfield  {author} {\bibinfo {author} {\bibfnamefont {R.}~\bibnamefont
  {Snellings}},\ }\href {\doibase 10.1088/1367-2630/13/5/055008} {\bibfield
  {journal} {\bibinfo  {journal} {New J. Phys.}\ }\textbf {\bibinfo {volume}
  {13}},\ \bibinfo {pages} {055008} (\bibinfo {year} {2011})},\ \Eprint
  {http://arxiv.org/abs/1102.3010}{arXiv:1102.3010 [nucl-ex]}\BibitemShut
  {NoStop}%
\bibitem [{\citenamefont {Heinz}\ and\ \citenamefont
  {Snellings}(2013)}]{Heinz:2013th}%
  \BibitemOpen
  \bibfield  {author} {\bibinfo {author} {\bibfnamefont {U.}~\bibnamefont
  {Heinz}}\ and\ \bibinfo {author} {\bibfnamefont {R.}~\bibnamefont
  {Snellings}},\ }\href {\doibase 10.1146/annurev-nucl-102212-170540}
  {\bibfield  {journal} {\bibinfo  {journal} {Ann. Rev. Nucl. Part. Sci.}\
  }\textbf {\bibinfo {volume} {63}},\ \bibinfo {pages} {123} (\bibinfo {year}
  {2013})},\ \Eprint {http://arxiv.org/abs/1301.2826}{arXiv:1301.2826
  [nucl-th]}\BibitemShut {NoStop}%
\bibitem [{\citenamefont {Aggarwal}(2021)}]{MadanAggarwal:2021}%
  \BibitemOpen
  \bibfield  {author} {\bibinfo {author} {\bibfnamefont {M.~M.}\ \bibnamefont
  {Aggarwal}},\ }in\ \href@noop {} {\emph {\bibinfo {booktitle} {Advances in
  Nuclear Physics}}},\ \bibinfo {editor} {edited by\ \bibinfo {editor}
  {\bibfnamefont {R.~K.}\ \bibnamefont {Puri}}, \bibinfo {editor}
  {\bibfnamefont {J.}~\bibnamefont {Aichelin}}, \bibinfo {editor}
  {\bibfnamefont {S.}~\bibnamefont {Gautam}}, \ and\ \bibinfo {editor}
  {\bibfnamefont {R.}~\bibnamefont {Kumar}}}\ (\bibinfo  {publisher} {Springer
  Singapore},\ \bibinfo {address} {Singapore},\ \bibinfo {year} {2021})\ pp.\
  \bibinfo {pages} {161--188}\BibitemShut {NoStop}%
\bibitem [{\citenamefont {Ollitrault}(2023)}]{Ollitrault:2023wjk}%
  \BibitemOpen
  \bibfield  {author} {\bibinfo {author} {\bibfnamefont {J.-Y.}\ \bibnamefont
  {Ollitrault}},\ }\href@noop {} {\  (\bibinfo {year} {2023})},\ \Eprint
  {http://arxiv.org/abs/2308.11674}{arXiv:2308.11674 [nucl-ex]}\BibitemShut
  {NoStop}%
\bibitem [{\citenamefont {Molnar}\ and\ \citenamefont
  {Sun}(2013)}]{Molnar:2013eqa}%
  \BibitemOpen
  \bibfield  {author} {\bibinfo {author} {\bibfnamefont {D.}~\bibnamefont
  {Molnar}}\ and\ \bibinfo {author} {\bibfnamefont {D.}~\bibnamefont {Sun}},\
  }\href@noop {} {\  (\bibinfo {year} {2013})},\ \Eprint
  {http://arxiv.org/abs/1305.1046}{arXiv:1305.1046 [nucl-th]}\BibitemShut
  {NoStop}%
\bibitem [{\citenamefont {Noronha-Hostler}\ \emph {et~al.}(2016)\citenamefont
  {Noronha-Hostler}, \citenamefont {Betz}, \citenamefont {Noronha},\ and\
  \citenamefont {Gyulassy}}]{Noronha-Hostler:2016eow}%
  \BibitemOpen
  \bibfield  {author} {\bibinfo {author} {\bibfnamefont {J.}~\bibnamefont
  {Noronha-Hostler}}, \bibinfo {author} {\bibfnamefont {B.}~\bibnamefont
  {Betz}}, \bibinfo {author} {\bibfnamefont {J.}~\bibnamefont {Noronha}}, \
  and\ \bibinfo {author} {\bibfnamefont {M.}~\bibnamefont {Gyulassy}},\ }\href
  {\doibase 10.1103/PhysRevLett.116.252301} {\bibfield  {journal} {\bibinfo
  {journal} {Phys. Rev. Lett.}\ }\textbf {\bibinfo {volume} {116}},\ \bibinfo
  {pages} {252301} (\bibinfo {year} {2016})},\ \Eprint
  {http://arxiv.org/abs/1602.03788}{arXiv:1602.03788 [nucl-th]}\BibitemShut
  {NoStop}%
\bibitem [{\citenamefont {Zhang}\ and\ \citenamefont
  {Liao}(2013)}]{Zhang:2013oca}%
  \BibitemOpen
  \bibfield  {author} {\bibinfo {author} {\bibfnamefont {X.}~\bibnamefont
  {Zhang}}\ and\ \bibinfo {author} {\bibfnamefont {J.}~\bibnamefont {Liao}},\
  }\href@noop {} {\  (\bibinfo {year} {2013})},\ \Eprint
  {http://arxiv.org/abs/1311.5463}{arXiv:1311.5463 [nucl-th]}\BibitemShut
  {NoStop}%
\bibitem [{\citenamefont {Liao}\ and\ \citenamefont
  {Shuryak}(2009)}]{Liao:2008dk}%
  \BibitemOpen
  \bibfield  {author} {\bibinfo {author} {\bibfnamefont {J.}~\bibnamefont
  {Liao}}\ and\ \bibinfo {author} {\bibfnamefont {E.}~\bibnamefont {Shuryak}},\
  }\href {\doibase 10.1103/PhysRevLett.102.202302} {\bibfield  {journal}
  {\bibinfo  {journal} {Phys. Rev. Lett.}\ }\textbf {\bibinfo {volume} {102}},\
  \bibinfo {pages} {202302} (\bibinfo {year} {2009})},\ \Eprint
  {http://arxiv.org/abs/0810.4116}{arXiv:0810.4116 [nucl-th]}\BibitemShut
  {NoStop}%
\bibitem [{\citenamefont {Kopeliovich}\ \emph {et~al.}(2012)\citenamefont
  {Kopeliovich}, \citenamefont {Nemchik}, \citenamefont {Potashnikova},\ and\
  \citenamefont {Schmidt}}]{Kopeliovich:2012sc}%
  \BibitemOpen
  \bibfield  {author} {\bibinfo {author} {\bibfnamefont {B.~Z.}\ \bibnamefont
  {Kopeliovich}}, \bibinfo {author} {\bibfnamefont {J.}~\bibnamefont
  {Nemchik}}, \bibinfo {author} {\bibfnamefont {I.~K.}\ \bibnamefont
  {Potashnikova}}, \ and\ \bibinfo {author} {\bibfnamefont {I.}~\bibnamefont
  {Schmidt}},\ }\href {\doibase 10.1103/PhysRevC.86.054904} {\bibfield
  {journal} {\bibinfo  {journal} {Phys. Rev. C}\ }\textbf {\bibinfo {volume}
  {86}},\ \bibinfo {pages} {054904} (\bibinfo {year} {2012})},\ \Eprint
  {http://arxiv.org/abs/1208.4951}{arXiv:1208.4951 [hep-ph]}\BibitemShut
  {NoStop}%
\bibitem [{\citenamefont {Cao}\ \emph {et~al.}(2017)\citenamefont {Cao},
  \citenamefont {Pang}, \citenamefont {Luo}, \citenamefont {He}, \citenamefont
  {Qin},\ and\ \citenamefont {Wang}}]{Cao:2017umt}%
  \BibitemOpen
  \bibfield  {author} {\bibinfo {author} {\bibfnamefont {S.}~\bibnamefont
  {Cao}}, \bibinfo {author} {\bibfnamefont {L.-G.}\ \bibnamefont {Pang}},
  \bibinfo {author} {\bibfnamefont {T.}~\bibnamefont {Luo}}, \bibinfo {author}
  {\bibfnamefont {Y.}~\bibnamefont {He}}, \bibinfo {author} {\bibfnamefont
  {G.-Y.}\ \bibnamefont {Qin}}, \ and\ \bibinfo {author} {\bibfnamefont
  {X.-N.}\ \bibnamefont {Wang}},\ }\href {\doibase
  10.1016/j.nuclphysbps.2017.05.048} {\bibfield  {journal} {\bibinfo  {journal}
  {Nucl. Part. Phys. Proc.}\ }\textbf {\bibinfo {volume} {289-290}},\ \bibinfo
  {pages} {217} (\bibinfo {year} {2017})}\BibitemShut {NoStop}%
\bibitem [{\citenamefont {Andres}\ \emph {et~al.}(2020)\citenamefont {Andres},
  \citenamefont {Armesto}, \citenamefont {Niemi}, \citenamefont {Paatelainen},\
  and\ \citenamefont {Salgado}}]{Andres:2019eus}%
  \BibitemOpen
  \bibfield  {author} {\bibinfo {author} {\bibfnamefont {C.}~\bibnamefont
  {Andres}}, \bibinfo {author} {\bibfnamefont {N.}~\bibnamefont {Armesto}},
  \bibinfo {author} {\bibfnamefont {H.}~\bibnamefont {Niemi}}, \bibinfo
  {author} {\bibfnamefont {R.}~\bibnamefont {Paatelainen}}, \ and\ \bibinfo
  {author} {\bibfnamefont {C.~A.}\ \bibnamefont {Salgado}},\ }\href {\doibase
  10.1016/j.physletb.2020.135318} {\bibfield  {journal} {\bibinfo  {journal}
  {Phys. Lett. B}\ }\textbf {\bibinfo {volume} {803}},\ \bibinfo {pages}
  {135318} (\bibinfo {year} {2020})},\ \Eprint
  {http://arxiv.org/abs/1902.03231}{arXiv:1902.03231 [hep-ph]}\BibitemShut
  {NoStop}%
\bibitem [{\citenamefont {Romatschke}\ and\ \citenamefont
  {Romatschke}(2007)}]{Romatschke:2007mq}%
  \BibitemOpen
  \bibfield  {author} {\bibinfo {author} {\bibfnamefont {P.}~\bibnamefont
  {Romatschke}}\ and\ \bibinfo {author} {\bibfnamefont {U.}~\bibnamefont
  {Romatschke}},\ }\href {\doibase 10.1103/PhysRevLett.99.172301} {\bibfield
  {journal} {\bibinfo  {journal} {Phys. Rev. Lett.}\ }\textbf {\bibinfo
  {volume} {99}},\ \bibinfo {pages} {172301} (\bibinfo {year} {2007})},\
  \Eprint {http://arxiv.org/abs/0706.1522}{arXiv:0706.1522
  [nucl-th]}\BibitemShut {NoStop}%
\bibitem [{\citenamefont {Song}\ \emph {et~al.}(2011)\citenamefont {Song},
  \citenamefont {Bass}, \citenamefont {Heinz}, \citenamefont {Hirano},\ and\
  \citenamefont {Shen}}]{Song:2010mg}%
  \BibitemOpen
  \bibfield  {author} {\bibinfo {author} {\bibfnamefont {H.}~\bibnamefont
  {Song}}, \bibinfo {author} {\bibfnamefont {S.~A.}\ \bibnamefont {Bass}},
  \bibinfo {author} {\bibfnamefont {U.}~\bibnamefont {Heinz}}, \bibinfo
  {author} {\bibfnamefont {T.}~\bibnamefont {Hirano}}, \ and\ \bibinfo {author}
  {\bibfnamefont {C.}~\bibnamefont {Shen}},\ }\href {\doibase
  10.1103/PhysRevLett.106.192301} {\bibfield  {journal} {\bibinfo  {journal}
  {Phys. Rev. Lett.}\ }\textbf {\bibinfo {volume} {106}},\ \bibinfo {pages}
  {192301} (\bibinfo {year} {2011})},\ \bibinfo {note} {[Erratum:
  Phys.Rev.Lett. 109, 139904 (2012)]},\ \Eprint
  {http://arxiv.org/abs/1011.2783}{arXiv:1011.2783 [nucl-th]}\BibitemShut
  {NoStop}%
\bibitem [{\citenamefont {Gale}\ \emph
  {et~al.}(2013{\natexlab{a}})\citenamefont {Gale}, \citenamefont {Jeon},
  \citenamefont {Schenke}, \citenamefont {Tribedy},\ and\ \citenamefont
  {Venugopalan}}]{Gale:2012rq}%
  \BibitemOpen
  \bibfield  {author} {\bibinfo {author} {\bibfnamefont {C.}~\bibnamefont
  {Gale}}, \bibinfo {author} {\bibfnamefont {S.}~\bibnamefont {Jeon}}, \bibinfo
  {author} {\bibfnamefont {B.}~\bibnamefont {Schenke}}, \bibinfo {author}
  {\bibfnamefont {P.}~\bibnamefont {Tribedy}}, \ and\ \bibinfo {author}
  {\bibfnamefont {R.}~\bibnamefont {Venugopalan}},\ }\href {\doibase
  10.1103/PhysRevLett.110.012302} {\bibfield  {journal} {\bibinfo  {journal}
  {Phys. Rev. Lett.}\ }\textbf {\bibinfo {volume} {110}},\ \bibinfo {pages}
  {012302} (\bibinfo {year} {2013}{\natexlab{a}})},\ \Eprint
  {http://arxiv.org/abs/1209.6330}{arXiv:1209.6330 [nucl-th]}\BibitemShut
  {NoStop}%
\bibitem [{\citenamefont {Huovinen}(2013)}]{Huovinen:2013wma}%
  \BibitemOpen
  \bibfield  {author} {\bibinfo {author} {\bibfnamefont {P.}~\bibnamefont
  {Huovinen}},\ }\href {\doibase 10.1142/S0218301313300294} {\bibfield
  {journal} {\bibinfo  {journal} {Int. J. Mod. Phys. E}\ }\textbf {\bibinfo
  {volume} {22}},\ \bibinfo {pages} {1330029} (\bibinfo {year} {2013})},\
  \Eprint {http://arxiv.org/abs/1311.1849}{arXiv:1311.1849
  [nucl-th]}\BibitemShut {NoStop}%
\bibitem [{\citenamefont {Gale}\ \emph
  {et~al.}(2013{\natexlab{b}})\citenamefont {Gale}, \citenamefont {Jeon},\ and\
  \citenamefont {Schenke}}]{Gale:2013da}%
  \BibitemOpen
  \bibfield  {author} {\bibinfo {author} {\bibfnamefont {C.}~\bibnamefont
  {Gale}}, \bibinfo {author} {\bibfnamefont {S.}~\bibnamefont {Jeon}}, \ and\
  \bibinfo {author} {\bibfnamefont {B.}~\bibnamefont {Schenke}},\ }\href
  {\doibase 10.1142/S0217751X13400113} {\bibfield  {journal} {\bibinfo
  {journal} {Int. J. Mod. Phys. A}\ }\textbf {\bibinfo {volume} {28}},\
  \bibinfo {pages} {1340011} (\bibinfo {year} {2013}{\natexlab{b}})},\ \Eprint
  {http://arxiv.org/abs/1301.5893}{arXiv:1301.5893 [nucl-th]}\BibitemShut
  {NoStop}%
\bibitem [{\citenamefont {Niemi}\ \emph {et~al.}(2016)\citenamefont {Niemi},
  \citenamefont {Eskola},\ and\ \citenamefont {Paatelainen}}]{Niemi:2015qia}%
  \BibitemOpen
  \bibfield  {author} {\bibinfo {author} {\bibfnamefont {H.}~\bibnamefont
  {Niemi}}, \bibinfo {author} {\bibfnamefont {K.~J.}\ \bibnamefont {Eskola}}, \
  and\ \bibinfo {author} {\bibfnamefont {R.}~\bibnamefont {Paatelainen}},\
  }\href {\doibase 10.1103/PhysRevC.93.024907} {\bibfield  {journal} {\bibinfo
  {journal} {Phys. Rev. C}\ }\textbf {\bibinfo {volume} {93}},\ \bibinfo
  {pages} {024907} (\bibinfo {year} {2016})},\ \Eprint
  {http://arxiv.org/abs/1505.02677}{arXiv:1505.02677 [hep-ph]}\BibitemShut
  {NoStop}%
\bibitem [{\citenamefont {Bernhard}\ \emph {et~al.}(2016)\citenamefont
  {Bernhard}, \citenamefont {Moreland}, \citenamefont {Bass}, \citenamefont
  {Liu},\ and\ \citenamefont {Heinz}}]{Bernhard:2016tnd}%
  \BibitemOpen
  \bibfield  {author} {\bibinfo {author} {\bibfnamefont {J.~E.}\ \bibnamefont
  {Bernhard}}, \bibinfo {author} {\bibfnamefont {J.~S.}\ \bibnamefont
  {Moreland}}, \bibinfo {author} {\bibfnamefont {S.~A.}\ \bibnamefont {Bass}},
  \bibinfo {author} {\bibfnamefont {J.}~\bibnamefont {Liu}}, \ and\ \bibinfo
  {author} {\bibfnamefont {U.}~\bibnamefont {Heinz}},\ }\href {\doibase
  10.1103/PhysRevC.94.024907} {\bibfield  {journal} {\bibinfo  {journal} {Phys.
  Rev. C}\ }\textbf {\bibinfo {volume} {94}},\ \bibinfo {pages} {024907}
  (\bibinfo {year} {2016})},\ \Eprint
  {http://arxiv.org/abs/1605.03954}{arXiv:1605.03954 [nucl-th]}\BibitemShut
  {NoStop}%
\bibitem [{\citenamefont {McDonald}\ \emph {et~al.}(2017)\citenamefont
  {McDonald}, \citenamefont {Shen}, \citenamefont {Fillion-Gourdeau},
  \citenamefont {Jeon},\ and\ \citenamefont {Gale}}]{McDonald:2016vlt}%
  \BibitemOpen
  \bibfield  {author} {\bibinfo {author} {\bibfnamefont {S.}~\bibnamefont
  {McDonald}}, \bibinfo {author} {\bibfnamefont {C.}~\bibnamefont {Shen}},
  \bibinfo {author} {\bibfnamefont {F.}~\bibnamefont {Fillion-Gourdeau}},
  \bibinfo {author} {\bibfnamefont {S.}~\bibnamefont {Jeon}}, \ and\ \bibinfo
  {author} {\bibfnamefont {C.}~\bibnamefont {Gale}},\ }\href {\doibase
  10.1103/PhysRevC.95.064913} {\bibfield  {journal} {\bibinfo  {journal} {Phys.
  Rev. C}\ }\textbf {\bibinfo {volume} {95}},\ \bibinfo {pages} {064913}
  (\bibinfo {year} {2017})},\ \Eprint
  {http://arxiv.org/abs/1609.02958}{arXiv:1609.02958 [hep-ph]}\BibitemShut
  {NoStop}%
\bibitem [{\citenamefont {Zhao}\ \emph {et~al.}(2017)\citenamefont {Zhao},
  \citenamefont {Xu},\ and\ \citenamefont {Song}}]{Zhao:2017yhj}%
  \BibitemOpen
  \bibfield  {author} {\bibinfo {author} {\bibfnamefont {W.}~\bibnamefont
  {Zhao}}, \bibinfo {author} {\bibfnamefont {H.-j.}\ \bibnamefont {Xu}}, \ and\
  \bibinfo {author} {\bibfnamefont {H.}~\bibnamefont {Song}},\ }\href {\doibase
  10.1140/epjc/s10052-017-5186-x} {\bibfield  {journal} {\bibinfo  {journal}
  {Eur. Phys. J. C}\ }\textbf {\bibinfo {volume} {77}},\ \bibinfo {pages} {645}
  (\bibinfo {year} {2017})},\ \Eprint
  {http://arxiv.org/abs/1703.10792}{arXiv:1703.10792 [nucl-th]}\BibitemShut
  {NoStop}%
\bibitem [{\citenamefont {Wang}\ and\ \citenamefont
  {Gyulassy}(1992)}]{Wang:1992qdg}%
  \BibitemOpen
  \bibfield  {author} {\bibinfo {author} {\bibfnamefont {X.-N.}\ \bibnamefont
  {Wang}}\ and\ \bibinfo {author} {\bibfnamefont {M.}~\bibnamefont
  {Gyulassy}},\ }\href {\doibase 10.1103/PhysRevLett.68.1480} {\bibfield
  {journal} {\bibinfo  {journal} {Phys. Rev. Lett.}\ }\textbf {\bibinfo
  {volume} {68}},\ \bibinfo {pages} {1480} (\bibinfo {year}
  {1992})}\BibitemShut {NoStop}%
\bibitem [{\citenamefont {Wang}(2000)}]{Wang:1998ww}%
  \BibitemOpen
  \bibfield  {author} {\bibinfo {author} {\bibfnamefont {X.-N.}\ \bibnamefont
  {Wang}},\ }\href {\doibase 10.1103/PhysRevC.61.064910} {\bibfield  {journal}
  {\bibinfo  {journal} {Phys. Rev. C}\ }\textbf {\bibinfo {volume} {61}},\
  \bibinfo {pages} {064910} (\bibinfo {year} {2000})},\ \Eprint
  {http://arxiv.org/abs/nucl-th/9812021}{arXiv:nucl-th/9812021}\BibitemShut
  {NoStop}%
\bibitem [{\citenamefont {Vitev}\ and\ \citenamefont
  {Gyulassy}(2002)}]{Vitev:2002pf}%
  \BibitemOpen
  \bibfield  {author} {\bibinfo {author} {\bibfnamefont {I.}~\bibnamefont
  {Vitev}}\ and\ \bibinfo {author} {\bibfnamefont {M.}~\bibnamefont
  {Gyulassy}},\ }\href {\doibase 10.1103/PhysRevLett.89.252301} {\bibfield
  {journal} {\bibinfo  {journal} {Phys. Rev. Lett.}\ }\textbf {\bibinfo
  {volume} {89}},\ \bibinfo {pages} {252301} (\bibinfo {year} {2002})},\
  \Eprint
  {http://arxiv.org/abs/hep-ph/0209161}{arXiv:hep-ph/0209161}\BibitemShut
  {NoStop}%
\bibitem [{\citenamefont {Wang}\ and\ \citenamefont
  {Wang}(2002)}]{Wang:2002ri}%
  \BibitemOpen
  \bibfield  {author} {\bibinfo {author} {\bibfnamefont {E.}~\bibnamefont
  {Wang}}\ and\ \bibinfo {author} {\bibfnamefont {X.-N.}\ \bibnamefont
  {Wang}},\ }\href {\doibase 10.1103/PhysRevLett.89.162301} {\bibfield
  {journal} {\bibinfo  {journal} {Phys. Rev. Lett.}\ }\textbf {\bibinfo
  {volume} {89}},\ \bibinfo {pages} {162301} (\bibinfo {year} {2002})},\
  \Eprint
  {http://arxiv.org/abs/hep-ph/0202105}{arXiv:hep-ph/0202105}\BibitemShut
  {NoStop}%
\bibitem [{\citenamefont {Eskola}\ \emph {et~al.}(2005)\citenamefont {Eskola},
  \citenamefont {Honkanen}, \citenamefont {Salgado},\ and\ \citenamefont
  {Wiedemann}}]{Eskola:2004cr}%
  \BibitemOpen
  \bibfield  {author} {\bibinfo {author} {\bibfnamefont {K.~J.}\ \bibnamefont
  {Eskola}}, \bibinfo {author} {\bibfnamefont {H.}~\bibnamefont {Honkanen}},
  \bibinfo {author} {\bibfnamefont {C.~A.}\ \bibnamefont {Salgado}}, \ and\
  \bibinfo {author} {\bibfnamefont {U.~A.}\ \bibnamefont {Wiedemann}},\ }\href
  {\doibase 10.1016/j.nuclphysa.2004.09.070} {\bibfield  {journal} {\bibinfo
  {journal} {Nucl. Phys. A}\ }\textbf {\bibinfo {volume} {747}},\ \bibinfo
  {pages} {511} (\bibinfo {year} {2005})},\ \Eprint
  {http://arxiv.org/abs/hep-ph/0406319}{arXiv:hep-ph/0406319}\BibitemShut
  {NoStop}%
\bibitem [{\citenamefont {Qin}\ \emph {et~al.}(2008)\citenamefont {Qin},
  \citenamefont {Ruppert}, \citenamefont {Gale}, \citenamefont {Jeon},
  \citenamefont {Moore},\ and\ \citenamefont {Mustafa}}]{Qin:2007rn}%
  \BibitemOpen
  \bibfield  {author} {\bibinfo {author} {\bibfnamefont {G.-Y.}\ \bibnamefont
  {Qin}}, \bibinfo {author} {\bibfnamefont {J.}~\bibnamefont {Ruppert}},
  \bibinfo {author} {\bibfnamefont {C.}~\bibnamefont {Gale}}, \bibinfo {author}
  {\bibfnamefont {S.}~\bibnamefont {Jeon}}, \bibinfo {author} {\bibfnamefont
  {G.~D.}\ \bibnamefont {Moore}}, \ and\ \bibinfo {author} {\bibfnamefont
  {M.~G.}\ \bibnamefont {Mustafa}},\ }\href {\doibase
  10.1103/PhysRevLett.100.072301} {\bibfield  {journal} {\bibinfo  {journal}
  {Phys. Rev. Lett.}\ }\textbf {\bibinfo {volume} {100}},\ \bibinfo {pages}
  {072301} (\bibinfo {year} {2008})},\ \Eprint
  {http://arxiv.org/abs/0710.0605}{arXiv:0710.0605 [hep-ph]}\BibitemShut
  {NoStop}%
\bibitem [{\citenamefont {Schenke}\ \emph {et~al.}(2009)\citenamefont
  {Schenke}, \citenamefont {Gale},\ and\ \citenamefont
  {Jeon}}]{Schenke:2009gb}%
  \BibitemOpen
  \bibfield  {author} {\bibinfo {author} {\bibfnamefont {B.}~\bibnamefont
  {Schenke}}, \bibinfo {author} {\bibfnamefont {C.}~\bibnamefont {Gale}}, \
  and\ \bibinfo {author} {\bibfnamefont {S.}~\bibnamefont {Jeon}},\ }\href
  {\doibase 10.1103/PhysRevC.80.054913} {\bibfield  {journal} {\bibinfo
  {journal} {Phys. Rev. C}\ }\textbf {\bibinfo {volume} {80}},\ \bibinfo
  {pages} {054913} (\bibinfo {year} {2009})},\ \Eprint
  {http://arxiv.org/abs/0909.2037}{arXiv:0909.2037 [hep-ph]}\BibitemShut
  {NoStop}%
\bibitem [{\citenamefont {Chen}\ \emph {et~al.}(2011)\citenamefont {Chen},
  \citenamefont {Hirano}, \citenamefont {Wang}, \citenamefont {Wang},\ and\
  \citenamefont {Zhang}}]{Chen:2011vt}%
  \BibitemOpen
  \bibfield  {author} {\bibinfo {author} {\bibfnamefont {X.-F.}\ \bibnamefont
  {Chen}}, \bibinfo {author} {\bibfnamefont {T.}~\bibnamefont {Hirano}},
  \bibinfo {author} {\bibfnamefont {E.}~\bibnamefont {Wang}}, \bibinfo {author}
  {\bibfnamefont {X.-N.}\ \bibnamefont {Wang}}, \ and\ \bibinfo {author}
  {\bibfnamefont {H.}~\bibnamefont {Zhang}},\ }\href {\doibase
  10.1103/PhysRevC.84.034902} {\bibfield  {journal} {\bibinfo  {journal} {Phys.
  Rev. C}\ }\textbf {\bibinfo {volume} {84}},\ \bibinfo {pages} {034902}
  (\bibinfo {year} {2011})},\ \Eprint
  {http://arxiv.org/abs/1102.5614}{arXiv:1102.5614 [nucl-th]}\BibitemShut
  {NoStop}%
\bibitem [{\citenamefont {Majumder}\ and\ \citenamefont
  {Shen}(2012)}]{Majumder:2011uk}%
  \BibitemOpen
  \bibfield  {author} {\bibinfo {author} {\bibfnamefont {A.}~\bibnamefont
  {Majumder}}\ and\ \bibinfo {author} {\bibfnamefont {C.}~\bibnamefont
  {Shen}},\ }\href {\doibase 10.1103/PhysRevLett.109.202301} {\bibfield
  {journal} {\bibinfo  {journal} {Phys. Rev. Lett.}\ }\textbf {\bibinfo
  {volume} {109}},\ \bibinfo {pages} {202301} (\bibinfo {year} {2012})},\
  \Eprint {http://arxiv.org/abs/1103.0809}{arXiv:1103.0809
  [hep-ph]}\BibitemShut {NoStop}%
\bibitem [{\citenamefont {Buzzatti}\ and\ \citenamefont
  {Gyulassy}(2012)}]{Buzzatti:2011vt}%
  \BibitemOpen
  \bibfield  {author} {\bibinfo {author} {\bibfnamefont {A.}~\bibnamefont
  {Buzzatti}}\ and\ \bibinfo {author} {\bibfnamefont {M.}~\bibnamefont
  {Gyulassy}},\ }\href {\doibase 10.1103/PhysRevLett.108.022301} {\bibfield
  {journal} {\bibinfo  {journal} {Phys. Rev. Lett.}\ }\textbf {\bibinfo
  {volume} {108}},\ \bibinfo {pages} {022301} (\bibinfo {year} {2012})},\
  \Eprint {http://arxiv.org/abs/1106.3061}{arXiv:1106.3061
  [hep-ph]}\BibitemShut {NoStop}%
\bibitem [{\citenamefont {Zapp}\ \emph {et~al.}(2013)\citenamefont {Zapp},
  \citenamefont {Krauss},\ and\ \citenamefont {Wiedemann}}]{Zapp:2012ak}%
  \BibitemOpen
  \bibfield  {author} {\bibinfo {author} {\bibfnamefont {K.~C.}\ \bibnamefont
  {Zapp}}, \bibinfo {author} {\bibfnamefont {F.}~\bibnamefont {Krauss}}, \ and\
  \bibinfo {author} {\bibfnamefont {U.~A.}\ \bibnamefont {Wiedemann}},\ }\href
  {\doibase 10.1007/JHEP03(2013)080} {\bibfield  {journal} {\bibinfo  {journal}
  {JHEP}\ }\textbf {\bibinfo {volume} {03}},\ \bibinfo {pages} {080} (\bibinfo
  {year} {2013})},\ \Eprint {http://arxiv.org/abs/1212.1599}{arXiv:1212.1599
  [hep-ph]}\BibitemShut {NoStop}%
\bibitem [{\citenamefont {Gyulassy}\ \emph {et~al.}(2001)\citenamefont
  {Gyulassy}, \citenamefont {Vitev},\ and\ \citenamefont
  {Wang}}]{Gyulassy:2000gk}%
  \BibitemOpen
  \bibfield  {author} {\bibinfo {author} {\bibfnamefont {M.}~\bibnamefont
  {Gyulassy}}, \bibinfo {author} {\bibfnamefont {I.}~\bibnamefont {Vitev}}, \
  and\ \bibinfo {author} {\bibfnamefont {X.~N.}\ \bibnamefont {Wang}},\ }\href
  {\doibase 10.1103/PhysRevLett.86.2537} {\bibfield  {journal} {\bibinfo
  {journal} {Phys. Rev. Lett.}\ }\textbf {\bibinfo {volume} {86}},\ \bibinfo
  {pages} {2537} (\bibinfo {year} {2001})},\ \Eprint
  {http://arxiv.org/abs/nucl-th/0012092}{arXiv:nucl-th/0012092}\BibitemShut
  {NoStop}%
\bibitem [{\citenamefont {Kumar}\ \emph {et~al.}(2017)\citenamefont {Kumar},
  \citenamefont {Bianchi}, \citenamefont {Elledge}, \citenamefont {Majumder},
  \citenamefont {Qin},\ and\ \citenamefont {Shen}}]{Kumar:2017des}%
  \BibitemOpen
  \bibfield  {author} {\bibinfo {author} {\bibfnamefont {A.}~\bibnamefont
  {Kumar}}, \bibinfo {author} {\bibfnamefont {E.}~\bibnamefont {Bianchi}},
  \bibinfo {author} {\bibfnamefont {J.}~\bibnamefont {Elledge}}, \bibinfo
  {author} {\bibfnamefont {A.}~\bibnamefont {Majumder}}, \bibinfo {author}
  {\bibfnamefont {G.-Y.}\ \bibnamefont {Qin}}, \ and\ \bibinfo {author}
  {\bibfnamefont {C.}~\bibnamefont {Shen}},\ }\href {\doibase
  10.1016/j.nuclphysa.2017.05.015} {\bibfield  {journal} {\bibinfo  {journal}
  {Nucl. Phys. A}\ }\textbf {\bibinfo {volume} {967}},\ \bibinfo {pages} {536}
  (\bibinfo {year} {2017})},\ \Eprint
  {http://arxiv.org/abs/1706.07547}{arXiv:1706.07547 [nucl-th]}\BibitemShut
  {NoStop}%
\bibitem [{\citenamefont {Zigic}\ \emph {et~al.}(2020)\citenamefont {Zigic},
  \citenamefont {Ilic}, \citenamefont {Djordjevic},\ and\ \citenamefont
  {Djordjevic}}]{Zigic:2019sth}%
  \BibitemOpen
  \bibfield  {author} {\bibinfo {author} {\bibfnamefont {D.}~\bibnamefont
  {Zigic}}, \bibinfo {author} {\bibfnamefont {B.}~\bibnamefont {Ilic}},
  \bibinfo {author} {\bibfnamefont {M.}~\bibnamefont {Djordjevic}}, \ and\
  \bibinfo {author} {\bibfnamefont {M.}~\bibnamefont {Djordjevic}},\ }\href
  {\doibase 10.1103/PhysRevC.101.064909} {\bibfield  {journal} {\bibinfo
  {journal} {Phys. Rev. C}\ }\textbf {\bibinfo {volume} {101}},\ \bibinfo
  {pages} {064909} (\bibinfo {year} {2020})},\ \Eprint
  {http://arxiv.org/abs/1908.11866}{arXiv:1908.11866 [hep-ph]}\BibitemShut
  {NoStop}%
\bibitem [{\citenamefont {Mehtar-Tani}\ \emph {et~al.}(2013)\citenamefont
  {Mehtar-Tani}, \citenamefont {Milhano},\ and\ \citenamefont
  {Tywoniuk}}]{Mehtar-Tani:2013pia}%
  \BibitemOpen
  \bibfield  {author} {\bibinfo {author} {\bibfnamefont {Y.}~\bibnamefont
  {Mehtar-Tani}}, \bibinfo {author} {\bibfnamefont {J.~G.}\ \bibnamefont
  {Milhano}}, \ and\ \bibinfo {author} {\bibfnamefont {K.}~\bibnamefont
  {Tywoniuk}},\ }\href {\doibase 10.1142/S0217751X13400137} {\bibfield
  {journal} {\bibinfo  {journal} {Int. J. Mod. Phys. A}\ }\textbf {\bibinfo
  {volume} {28}},\ \bibinfo {pages} {1340013} (\bibinfo {year} {2013})},\
  \Eprint {http://arxiv.org/abs/1302.2579}{arXiv:1302.2579
  [hep-ph]}\BibitemShut {NoStop}%
\bibitem [{\citenamefont {Qin}\ and\ \citenamefont {Wang}(2015)}]{Qin:2015srf}%
  \BibitemOpen
  \bibfield  {author} {\bibinfo {author} {\bibfnamefont {G.-Y.}\ \bibnamefont
  {Qin}}\ and\ \bibinfo {author} {\bibfnamefont {X.-N.}\ \bibnamefont {Wang}},\
  }\href {\doibase 10.1142/S0218301315300143} {\bibfield  {journal} {\bibinfo
  {journal} {Int. J. Mod. Phys. E}\ }\textbf {\bibinfo {volume} {24}},\
  \bibinfo {pages} {1530014} (\bibinfo {year} {2015})},\ \Eprint
  {http://arxiv.org/abs/1511.00790}{arXiv:1511.00790 [hep-ph]}\BibitemShut
  {NoStop}%
\bibitem [{\citenamefont {Blaizot}\ and\ \citenamefont
  {Mehtar-Tani}(2015)}]{Blaizot:2015lma}%
  \BibitemOpen
  \bibfield  {author} {\bibinfo {author} {\bibfnamefont {J.-P.}\ \bibnamefont
  {Blaizot}}\ and\ \bibinfo {author} {\bibfnamefont {Y.}~\bibnamefont
  {Mehtar-Tani}},\ }\href {\doibase 10.1142/S021830131530012X} {\bibfield
  {journal} {\bibinfo  {journal} {Int. J. Mod. Phys. E}\ }\textbf {\bibinfo
  {volume} {24}},\ \bibinfo {pages} {1530012} (\bibinfo {year} {2015})},\
  \Eprint {http://arxiv.org/abs/1503.05958}{arXiv:1503.05958
  [hep-ph]}\BibitemShut {NoStop}%
\bibitem [{\citenamefont {Zhao}\ \emph {et~al.}(2022)\citenamefont {Zhao},
  \citenamefont {Ke}, \citenamefont {Chen}, \citenamefont {Luo},\ and\
  \citenamefont {Wang}}]{Zhao:2021vmu}%
  \BibitemOpen
  \bibfield  {author} {\bibinfo {author} {\bibfnamefont {W.}~\bibnamefont
  {Zhao}}, \bibinfo {author} {\bibfnamefont {W.}~\bibnamefont {Ke}}, \bibinfo
  {author} {\bibfnamefont {W.}~\bibnamefont {Chen}}, \bibinfo {author}
  {\bibfnamefont {T.}~\bibnamefont {Luo}}, \ and\ \bibinfo {author}
  {\bibfnamefont {X.-N.}\ \bibnamefont {Wang}},\ }\href {\doibase
  10.1103/PhysRevLett.128.022302} {\bibfield  {journal} {\bibinfo  {journal}
  {Phys. Rev. Lett.}\ }\textbf {\bibinfo {volume} {128}},\ \bibinfo {pages}
  {022302} (\bibinfo {year} {2022})},\ \Eprint
  {http://arxiv.org/abs/2103.14657}{arXiv:2103.14657 [hep-ph]}\BibitemShut
  {NoStop}%
\bibitem [{\citenamefont {Barreto}\ \emph {et~al.}(2022)\citenamefont
  {Barreto}, \citenamefont {Canedo}, \citenamefont {Munhoz}, \citenamefont
  {Noronha},\ and\ \citenamefont {Noronha-Hostler}}]{Barreto:2022ulg}%
  \BibitemOpen
  \bibfield  {author} {\bibinfo {author} {\bibfnamefont {L.}~\bibnamefont
  {Barreto}}, \bibinfo {author} {\bibfnamefont {F.~M.}\ \bibnamefont {Canedo}},
  \bibinfo {author} {\bibfnamefont {M.~G.}\ \bibnamefont {Munhoz}}, \bibinfo
  {author} {\bibfnamefont {J.}~\bibnamefont {Noronha}}, \ and\ \bibinfo
  {author} {\bibfnamefont {J.}~\bibnamefont {Noronha-Hostler}},\ }\href@noop {}
  {\  (\bibinfo {year} {2022})},\ \Eprint
  {http://arxiv.org/abs/2208.02061}{arXiv:2208.02061 [nucl-th]}\BibitemShut
  {NoStop}%
\bibitem [{\citenamefont {Stojku}\ \emph {et~al.}(2022)\citenamefont {Stojku},
  \citenamefont {Auvinen}, \citenamefont {Djordjevic}, \citenamefont
  {Huovinen},\ and\ \citenamefont {Djordjevic}}]{Stojku:2020wkh}%
  \BibitemOpen
  \bibfield  {author} {\bibinfo {author} {\bibfnamefont {S.}~\bibnamefont
  {Stojku}}, \bibinfo {author} {\bibfnamefont {J.}~\bibnamefont {Auvinen}},
  \bibinfo {author} {\bibfnamefont {M.}~\bibnamefont {Djordjevic}}, \bibinfo
  {author} {\bibfnamefont {P.}~\bibnamefont {Huovinen}}, \ and\ \bibinfo
  {author} {\bibfnamefont {M.}~\bibnamefont {Djordjevic}},\ }\href {\doibase
  10.1103/PhysRevC.105.L021901} {\bibfield  {journal} {\bibinfo  {journal}
  {Phys. Rev. C}\ }\textbf {\bibinfo {volume} {105}},\ \bibinfo {pages}
  {L021901} (\bibinfo {year} {2022})},\ \Eprint
  {http://arxiv.org/abs/2008.08987}{arXiv:2008.08987 [nucl-th]}\BibitemShut
  {NoStop}%
\bibitem [{\citenamefont {Florkowski}\ and\ \citenamefont
  {Ryblewski}(2016)}]{Florkowski:2016qig}%
  \BibitemOpen
  \bibfield  {author} {\bibinfo {author} {\bibfnamefont {W.}~\bibnamefont
  {Florkowski}}\ and\ \bibinfo {author} {\bibfnamefont {R.}~\bibnamefont
  {Ryblewski}},\ }\href {\doibase 10.1103/PhysRevC.93.064903} {\bibfield
  {journal} {\bibinfo  {journal} {Phys. Rev. C}\ }\textbf {\bibinfo {volume}
  {93}},\ \bibinfo {pages} {064903} (\bibinfo {year} {2016})},\ \Eprint
  {http://arxiv.org/abs/1603.01704}{arXiv:1603.01704 [nucl-th]}\BibitemShut
  {NoStop}%
\bibitem [{\citenamefont {Tripathy}\ \emph {et~al.}(2016)\citenamefont
  {Tripathy}, \citenamefont {Bhattacharyya}, \citenamefont {Garg},
  \citenamefont {Kumar}, \citenamefont {Sahoo},\ and\ \citenamefont
  {Cleymans}}]{Tripathy_2016}%
  \BibitemOpen
  \bibfield  {author} {\bibinfo {author} {\bibfnamefont {S.}~\bibnamefont
  {Tripathy}}, \bibinfo {author} {\bibfnamefont {T.}~\bibnamefont
  {Bhattacharyya}}, \bibinfo {author} {\bibfnamefont {P.}~\bibnamefont {Garg}},
  \bibinfo {author} {\bibfnamefont {P.}~\bibnamefont {Kumar}}, \bibinfo
  {author} {\bibfnamefont {R.}~\bibnamefont {Sahoo}}, \ and\ \bibinfo {author}
  {\bibfnamefont {J.}~\bibnamefont {Cleymans}},\ }\href {\doibase
  10.1140/epja/i2016-16289-4} {\bibfield  {journal} {\bibinfo  {journal} {The
  European Physical Journal A}\ }\textbf {\bibinfo {volume} {52}} (\bibinfo
  {year} {2016}),\ 10.1140/epja/i2016-16289-4}\BibitemShut {NoStop}%
\bibitem [{\citenamefont {Tripathy}\ \emph {et~al.}(2017)\citenamefont
  {Tripathy}, \citenamefont {Khuntia}, \citenamefont {Tiwari},\ and\
  \citenamefont {Sahoo}}]{Tripathy:2017kwb}%
  \BibitemOpen
  \bibfield  {author} {\bibinfo {author} {\bibfnamefont {S.}~\bibnamefont
  {Tripathy}}, \bibinfo {author} {\bibfnamefont {A.}~\bibnamefont {Khuntia}},
  \bibinfo {author} {\bibfnamefont {S.~K.}\ \bibnamefont {Tiwari}}, \ and\
  \bibinfo {author} {\bibfnamefont {R.}~\bibnamefont {Sahoo}},\ }\href
  {\doibase 10.1140/epja/i2017-12283-8} {\bibfield  {journal} {\bibinfo
  {journal} {Eur. Phys. J. A}\ }\textbf {\bibinfo {volume} {53}},\ \bibinfo
  {pages} {99} (\bibinfo {year} {2017})},\ \Eprint
  {http://arxiv.org/abs/1703.02416}{arXiv:1703.02416 [nucl-th]}\BibitemShut
  {NoStop}%
\bibitem [{\citenamefont {Qiao}\ \emph {et~al.}(2020)\citenamefont {Qiao},
  \citenamefont {Che}, \citenamefont {Gu}, \citenamefont {Zheng},\ and\
  \citenamefont {Zhang}}]{Qiao_2020}%
  \BibitemOpen
  \bibfield  {author} {\bibinfo {author} {\bibfnamefont {L.}~\bibnamefont
  {Qiao}}, \bibinfo {author} {\bibfnamefont {G.}~\bibnamefont {Che}}, \bibinfo
  {author} {\bibfnamefont {J.}~\bibnamefont {Gu}}, \bibinfo {author}
  {\bibfnamefont {H.}~\bibnamefont {Zheng}}, \ and\ \bibinfo {author}
  {\bibfnamefont {W.}~\bibnamefont {Zhang}},\ }\href {\doibase
  10.1088/1361-6471/ab8744} {\bibfield  {journal} {\bibinfo  {journal} {Journal
  of Physics G: Nuclear and Particle Physics}\ }\textbf {\bibinfo {volume}
  {47}},\ \bibinfo {pages} {075101} (\bibinfo {year} {2020})}\BibitemShut
  {NoStop}%
\bibitem [{\citenamefont {Moriggi}\ \emph {et~al.}(2021)\citenamefont
  {Moriggi}, \citenamefont {Peccini},\ and\ \citenamefont
  {Machado}}]{Moriggi:2020qla}%
  \BibitemOpen
  \bibfield  {author} {\bibinfo {author} {\bibfnamefont {L.~S.}\ \bibnamefont
  {Moriggi}}, \bibinfo {author} {\bibfnamefont {G.~M.}\ \bibnamefont
  {Peccini}}, \ and\ \bibinfo {author} {\bibfnamefont {M.~V.~T.}\ \bibnamefont
  {Machado}},\ }\href {\doibase 10.1103/PhysRevD.103.034025} {\bibfield
  {journal} {\bibinfo  {journal} {Phys. Rev. D}\ }\textbf {\bibinfo {volume}
  {103}},\ \bibinfo {pages} {034025} (\bibinfo {year} {2021})},\ \Eprint
  {http://arxiv.org/abs/2012.05388}{arXiv:2012.05388 [hep-ph]}\BibitemShut
  {NoStop}%
\bibitem [{\citenamefont {Moriggi}\ and\ \citenamefont
  {Machado}(2022)}]{Moriggi:2022xbg}%
  \BibitemOpen
  \bibfield  {author} {\bibinfo {author} {\bibfnamefont {L.}~\bibnamefont
  {Moriggi}}\ and\ \bibinfo {author} {\bibfnamefont {M.~V.~T.}\ \bibnamefont
  {Machado}},\ }\href {\doibase 10.3390/physics4030050} {\bibfield  {journal}
  {\bibinfo  {journal} {MDPI Physics}\ }\textbf {\bibinfo {volume} {4}},\
  \bibinfo {pages} {787} (\bibinfo {year} {2022})},\ \Eprint
  {http://arxiv.org/abs/2207.07794}{arXiv:2207.07794 [hep-ph]}\BibitemShut
  {NoStop}%
\bibitem [{\citenamefont {Moriggi}\ \emph {et~al.}(2020)\citenamefont
  {Moriggi}, \citenamefont {Peccini},\ and\ \citenamefont
  {Machado}}]{Moriggi:2020zbv}%
  \BibitemOpen
  \bibfield  {author} {\bibinfo {author} {\bibfnamefont {L.~S.}\ \bibnamefont
  {Moriggi}}, \bibinfo {author} {\bibfnamefont {G.~M.}\ \bibnamefont
  {Peccini}}, \ and\ \bibinfo {author} {\bibfnamefont {M.~V.~T.}\ \bibnamefont
  {Machado}},\ }\href {\doibase 10.1103/PhysRevD.102.034016} {\bibfield
  {journal} {\bibinfo  {journal} {Phys. Rev. D}\ }\textbf {\bibinfo {volume}
  {102}},\ \bibinfo {pages} {034016} (\bibinfo {year} {2020})},\ \Eprint
  {http://arxiv.org/abs/2005.07760}{arXiv:2005.07760 [hep-ph]}\BibitemShut
  {NoStop}%
\bibitem [{\citenamefont {Tripathy}\ \emph {et~al.}(2018)\citenamefont
  {Tripathy}, \citenamefont {Tiwari}, \citenamefont {Younus},\ and\
  \citenamefont {Sahoo}}]{Tripathy:2017nmo}%
  \BibitemOpen
  \bibfield  {author} {\bibinfo {author} {\bibfnamefont {S.}~\bibnamefont
  {Tripathy}}, \bibinfo {author} {\bibfnamefont {S.~K.}\ \bibnamefont
  {Tiwari}}, \bibinfo {author} {\bibfnamefont {M.}~\bibnamefont {Younus}}, \
  and\ \bibinfo {author} {\bibfnamefont {R.}~\bibnamefont {Sahoo}},\ }\href
  {\doibase 10.1140/epja/i2018-12461-2} {\bibfield  {journal} {\bibinfo
  {journal} {Eur. Phys. J. A}\ }\textbf {\bibinfo {volume} {54}},\ \bibinfo
  {pages} {38} (\bibinfo {year} {2018})},\ \Eprint
  {http://arxiv.org/abs/1709.06354}{arXiv:1709.06354 [hep-ph]}\BibitemShut
  {NoStop}%
\bibitem [{\citenamefont {Younus}\ \emph {et~al.}(2020)\citenamefont {Younus},
  \citenamefont {Tripathy}, \citenamefont {Tiwari},\ and\ \citenamefont
  {Sahoo}}]{Younus:2018mrk}%
  \BibitemOpen
  \bibfield  {author} {\bibinfo {author} {\bibfnamefont {M.}~\bibnamefont
  {Younus}}, \bibinfo {author} {\bibfnamefont {S.}~\bibnamefont {Tripathy}},
  \bibinfo {author} {\bibfnamefont {S.~K.}\ \bibnamefont {Tiwari}}, \ and\
  \bibinfo {author} {\bibfnamefont {R.}~\bibnamefont {Sahoo}},\ }\href
  {\doibase 10.1155/2020/4728649} {\bibfield  {journal} {\bibinfo  {journal}
  {Adv. High Energy Phys.}\ }\textbf {\bibinfo {volume} {2020}},\ \bibinfo
  {pages} {4728649} (\bibinfo {year} {2020})},\ \Eprint
  {http://arxiv.org/abs/1803.01578}{arXiv:1803.01578 [hep-ph]}\BibitemShut
  {NoStop}%
\bibitem [{\citenamefont {Huovinen}\ \emph {et~al.}(2001)\citenamefont
  {Huovinen}, \citenamefont {Kolb}, \citenamefont {Heinz}, \citenamefont
  {Ruuskanen},\ and\ \citenamefont {Voloshin}}]{Huovinen:2001cy}%
  \BibitemOpen
  \bibfield  {author} {\bibinfo {author} {\bibfnamefont {P.}~\bibnamefont
  {Huovinen}}, \bibinfo {author} {\bibfnamefont {P.~F.}\ \bibnamefont {Kolb}},
  \bibinfo {author} {\bibfnamefont {U.~W.}\ \bibnamefont {Heinz}}, \bibinfo
  {author} {\bibfnamefont {P.~V.}\ \bibnamefont {Ruuskanen}}, \ and\ \bibinfo
  {author} {\bibfnamefont {S.~A.}\ \bibnamefont {Voloshin}},\ }\href {\doibase
  10.1016/S0370-2693(01)00219-2} {\bibfield  {journal} {\bibinfo  {journal}
  {Phys. Lett. B}\ }\textbf {\bibinfo {volume} {503}},\ \bibinfo {pages} {58}
  (\bibinfo {year} {2001})},\ \Eprint
  {http://arxiv.org/abs/hep-ph/0101136}{arXiv:hep-ph/0101136}\BibitemShut
  {NoStop}%
\bibitem [{\citenamefont {Akhil}\ and\ \citenamefont
  {Tiwari}(2023)}]{Akhil:2023xpb}%
  \BibitemOpen
  \bibfield  {author} {\bibinfo {author} {\bibfnamefont {A.}~\bibnamefont
  {Akhil}}\ and\ \bibinfo {author} {\bibfnamefont {S.~K.}\ \bibnamefont
  {Tiwari}},\ }\href@noop {} {\  (\bibinfo {year} {2023})},\ \Eprint
  {http://arxiv.org/abs/2309.06128}{arXiv:2309.06128 [hep-ph]}\BibitemShut
  {NoStop}%
\bibitem [{\citenamefont {Glauber}(1955)}]{Glauber:1955qq}%
  \BibitemOpen
  \bibfield  {author} {\bibinfo {author} {\bibfnamefont {R.}~\bibnamefont
  {Glauber}},\ }\href {\doibase 10.1103/PhysRev.100.242} {\bibfield  {journal}
  {\bibinfo  {journal} {Phys. Rev.}\ }\textbf {\bibinfo {volume} {100}},\
  \bibinfo {pages} {242} (\bibinfo {year} {1955})}\BibitemShut {NoStop}%
\bibitem [{\citenamefont {Mueller}(1990)}]{Mueller:1989st}%
  \BibitemOpen
  \bibfield  {author} {\bibinfo {author} {\bibfnamefont {A.~H.}\ \bibnamefont
  {Mueller}},\ }\href {\doibase 10.1016/0550-3213(90)90173-B} {\bibfield
  {journal} {\bibinfo  {journal} {Nucl. Phys. B}\ }\textbf {\bibinfo {volume}
  {335}},\ \bibinfo {pages} {115} (\bibinfo {year} {1990})}\BibitemShut
  {NoStop}%
\bibitem [{\citenamefont {Acharya}\ \emph {et~al.}(2020)\citenamefont {Acharya}
  \emph {et~al.}}]{ALICE:2019hno}%
  \BibitemOpen
  \bibfield  {author} {\bibinfo {author} {\bibfnamefont {S.}~\bibnamefont
  {Acharya}} \emph {et~al.} (\bibinfo {collaboration} {ALICE}),\ }\href
  {\doibase 10.1103/PhysRevC.101.044907} {\bibfield  {journal} {\bibinfo
  {journal} {Phys. Rev. C}\ }\textbf {\bibinfo {volume} {101}},\ \bibinfo
  {pages} {044907} (\bibinfo {year} {2020})},\ \Eprint
  {http://arxiv.org/abs/1910.07678}{arXiv:1910.07678 [nucl-ex]}\BibitemShut
  {NoStop}%
\bibitem [{\citenamefont {Acharya}\ \emph
  {et~al.}(2021{\natexlab{a}})\citenamefont {Acharya} \emph
  {et~al.}}]{ALICE:2021lsv}%
  \BibitemOpen
  \bibfield  {author} {\bibinfo {author} {\bibfnamefont {S.}~\bibnamefont
  {Acharya}} \emph {et~al.} (\bibinfo {collaboration} {ALICE}),\ }\href
  {\doibase 10.1140/epjc/s10052-021-09304-4} {\bibfield  {journal} {\bibinfo
  {journal} {Eur. Phys. J. C}\ }\textbf {\bibinfo {volume} {81}},\ \bibinfo
  {pages} {584} (\bibinfo {year} {2021}{\natexlab{a}})},\ \Eprint
  {http://arxiv.org/abs/2101.03100}{arXiv:2101.03100 [nucl-ex]}\BibitemShut
  {NoStop}%
\bibitem [{\citenamefont {Iancu}\ and\ \citenamefont
  {Rezaeian}(2017)}]{Iancu:2017fzn}%
  \BibitemOpen
  \bibfield  {author} {\bibinfo {author} {\bibfnamefont {E.}~\bibnamefont
  {Iancu}}\ and\ \bibinfo {author} {\bibfnamefont {A.~H.}\ \bibnamefont
  {Rezaeian}},\ }\href {\doibase 10.1103/PhysRevD.95.094003} {\bibfield
  {journal} {\bibinfo  {journal} {Phys. Rev. D}\ }\textbf {\bibinfo {volume}
  {95}},\ \bibinfo {pages} {094003} (\bibinfo {year} {2017})},\ \Eprint
  {http://arxiv.org/abs/1702.03943}{arXiv:1702.03943 [hep-ph]}\BibitemShut
  {NoStop}%
\bibitem [{\citenamefont {Kopeliovich}\ \emph {et~al.}(2021)\citenamefont
  {Kopeliovich}, \citenamefont {Krelina},\ and\ \citenamefont
  {Nemchik}}]{Kopeliovich:2021dgx}%
  \BibitemOpen
  \bibfield  {author} {\bibinfo {author} {\bibfnamefont {B.~Z.}\ \bibnamefont
  {Kopeliovich}}, \bibinfo {author} {\bibfnamefont {M.}~\bibnamefont
  {Krelina}}, \ and\ \bibinfo {author} {\bibfnamefont {J.}~\bibnamefont
  {Nemchik}},\ }\href {\doibase 10.1103/PhysRevD.103.094027} {\bibfield
  {journal} {\bibinfo  {journal} {Phys. Rev. D}\ }\textbf {\bibinfo {volume}
  {103}},\ \bibinfo {pages} {094027} (\bibinfo {year} {2021})},\ \Eprint
  {http://arxiv.org/abs/2102.06106}{arXiv:2102.06106 [hep-ph]}\BibitemShut
  {NoStop}%
\bibitem [{\citenamefont {Altinoluk}\ \emph {et~al.}(2016)\citenamefont
  {Altinoluk}, \citenamefont {Armesto}, \citenamefont {Beuf},\ and\
  \citenamefont {Rezaeian}}]{ALTINOLUK2016373}%
  \BibitemOpen
  \bibfield  {author} {\bibinfo {author} {\bibfnamefont {T.}~\bibnamefont
  {Altinoluk}}, \bibinfo {author} {\bibfnamefont {N.}~\bibnamefont {Armesto}},
  \bibinfo {author} {\bibfnamefont {G.}~\bibnamefont {Beuf}}, \ and\ \bibinfo
  {author} {\bibfnamefont {A.~H.}\ \bibnamefont {Rezaeian}},\ }\href {\doibase
  https://doi.org/10.1016/j.physletb.2016.05.032} {\bibfield  {journal}
  {\bibinfo  {journal} {Physics Letters B}\ }\textbf {\bibinfo {volume}
  {758}},\ \bibinfo {pages} {373} (\bibinfo {year} {2016})}\BibitemShut
  {NoStop}%
\bibitem [{\citenamefont {M\"antysaari}\ \emph {et~al.}(2019)\citenamefont
  {M\"antysaari}, \citenamefont {Mueller},\ and\ \citenamefont
  {Schenke}}]{PhysRevD.99.074004}%
  \BibitemOpen
  \bibfield  {author} {\bibinfo {author} {\bibfnamefont {H.}~\bibnamefont
  {M\"antysaari}}, \bibinfo {author} {\bibfnamefont {N.}~\bibnamefont
  {Mueller}}, \ and\ \bibinfo {author} {\bibfnamefont {B.}~\bibnamefont
  {Schenke}},\ }\href {\doibase 10.1103/PhysRevD.99.074004} {\bibfield
  {journal} {\bibinfo  {journal} {Phys. Rev. D}\ }\textbf {\bibinfo {volume}
  {99}},\ \bibinfo {pages} {074004} (\bibinfo {year} {2019})}\BibitemShut
  {NoStop}%
\bibitem [{\citenamefont {Salazar}\ and\ \citenamefont
  {Schenke}(2019)}]{PhysRevD.100.034007}%
  \BibitemOpen
  \bibfield  {author} {\bibinfo {author} {\bibfnamefont {F.}~\bibnamefont
  {Salazar}}\ and\ \bibinfo {author} {\bibfnamefont {B.}~\bibnamefont
  {Schenke}},\ }\href {\doibase 10.1103/PhysRevD.100.034007} {\bibfield
  {journal} {\bibinfo  {journal} {Phys. Rev. D}\ }\textbf {\bibinfo {volume}
  {100}},\ \bibinfo {pages} {034007} (\bibinfo {year} {2019})}\BibitemShut
  {NoStop}%
\bibitem [{\citenamefont {Golec-Biernat}\ and\ \citenamefont
  {Stasto}(2003)}]{Golec-Biernat:2003naj}%
  \BibitemOpen
  \bibfield  {author} {\bibinfo {author} {\bibfnamefont {K.~J.}\ \bibnamefont
  {Golec-Biernat}}\ and\ \bibinfo {author} {\bibfnamefont {A.~M.}\ \bibnamefont
  {Stasto}},\ }\href {\doibase 10.1016/j.nuclphysb.2003.07.011} {\bibfield
  {journal} {\bibinfo  {journal} {Nucl. Phys. B}\ }\textbf {\bibinfo {volume}
  {668}},\ \bibinfo {pages} {345} (\bibinfo {year} {2003})},\ \Eprint
  {http://arxiv.org/abs/hep-ph/0306279}{arXiv:hep-ph/0306279}\BibitemShut
  {NoStop}%
\bibitem [{\citenamefont {Berger}\ and\ \citenamefont
  {Sta\ifmmode~\acute{s}\else \'{s}\fi{}to}(2011)}]{PhysRevD.83.034015}%
  \BibitemOpen
  \bibfield  {author} {\bibinfo {author} {\bibfnamefont {J.}~\bibnamefont
  {Berger}}\ and\ \bibinfo {author} {\bibfnamefont {A.~M.}\ \bibnamefont
  {Sta\ifmmode~\acute{s}\else \'{s}\fi{}to}},\ }\href {\doibase
  10.1103/PhysRevD.83.034015} {\bibfield  {journal} {\bibinfo  {journal} {Phys.
  Rev. D}\ }\textbf {\bibinfo {volume} {83}},\ \bibinfo {pages} {034015}
  (\bibinfo {year} {2011})}\BibitemShut {NoStop}%
\bibitem [{\citenamefont {Schnedermann}\ \emph {et~al.}(1993)\citenamefont
  {Schnedermann}, \citenamefont {Sollfrank},\ and\ \citenamefont
  {Heinz}}]{PhysRevC.48.2462}%
  \BibitemOpen
  \bibfield  {author} {\bibinfo {author} {\bibfnamefont {E.}~\bibnamefont
  {Schnedermann}}, \bibinfo {author} {\bibfnamefont {J.}~\bibnamefont
  {Sollfrank}}, \ and\ \bibinfo {author} {\bibfnamefont {U.}~\bibnamefont
  {Heinz}},\ }\href {\doibase 10.1103/PhysRevC.48.2462} {\bibfield  {journal}
  {\bibinfo  {journal} {Phys. Rev. C}\ }\textbf {\bibinfo {volume} {48}},\
  \bibinfo {pages} {2462} (\bibinfo {year} {1993})}\BibitemShut {NoStop}%
\bibitem [{\citenamefont {Adler}\ \emph {et~al.}(2001)\citenamefont {Adler}
  \emph {et~al.}}]{STAR:2001ksn}%
  \BibitemOpen
  \bibfield  {author} {\bibinfo {author} {\bibfnamefont {C.}~\bibnamefont
  {Adler}} \emph {et~al.} (\bibinfo {collaboration} {STAR}),\ }\href {\doibase
  10.1103/PhysRevLett.87.182301} {\bibfield  {journal} {\bibinfo  {journal}
  {Phys. Rev. Lett.}\ }\textbf {\bibinfo {volume} {87}},\ \bibinfo {pages}
  {182301} (\bibinfo {year} {2001})},\ \Eprint
  {http://arxiv.org/abs/nucl-ex/0107003}{arXiv:nucl-ex/0107003}\BibitemShut
  {NoStop}%
\bibitem [{\citenamefont {Acharya}\ \emph {et~al.}(2018)\citenamefont {Acharya}
  \emph {et~al.}}]{ALICE:2018yph}%
  \BibitemOpen
  \bibfield  {author} {\bibinfo {author} {\bibfnamefont {S.}~\bibnamefont
  {Acharya}} \emph {et~al.} (\bibinfo {collaboration} {ALICE}),\ }\href
  {\doibase 10.1007/JHEP09(2018)006} {\bibfield  {journal} {\bibinfo  {journal}
  {JHEP}\ }\textbf {\bibinfo {volume} {09}},\ \bibinfo {pages} {006} (\bibinfo
  {year} {2018})},\ \Eprint {http://arxiv.org/abs/1805.04390}{arXiv:1805.04390
  [nucl-ex]}\BibitemShut {NoStop}%
\bibitem [{\citenamefont {Acharya}\ \emph
  {et~al.}(2021{\natexlab{b}})\citenamefont {Acharya} \emph
  {et~al.}}]{ALICE:2021ibz}%
  \BibitemOpen
  \bibfield  {author} {\bibinfo {author} {\bibfnamefont {S.}~\bibnamefont
  {Acharya}} \emph {et~al.} (\bibinfo {collaboration} {ALICE}),\ }\href
  {\doibase 10.1007/JHEP10(2021)152} {\bibfield  {journal} {\bibinfo  {journal}
  {JHEP}\ }\textbf {\bibinfo {volume} {10}},\ \bibinfo {pages} {152} (\bibinfo
  {year} {2021}{\natexlab{b}})},\ \Eprint
  {http://arxiv.org/abs/2107.10592}{arXiv:2107.10592 [nucl-ex]}\BibitemShut
  {NoStop}%
\bibitem [{\citenamefont {{De Vries}}\ \emph {et~al.}(1987)\citenamefont {{De
  Vries}}, \citenamefont {{De Jager}},\ and\ \citenamefont {{De
  Vries}}}]{DEVRIES1987495}%
  \BibitemOpen
  \bibfield  {author} {\bibinfo {author} {\bibfnamefont {H.}~\bibnamefont {{De
  Vries}}}, \bibinfo {author} {\bibfnamefont {C.}~\bibnamefont {{De Jager}}}, \
  and\ \bibinfo {author} {\bibfnamefont {C.}~\bibnamefont {{De Vries}}},\
  }\href {\doibase https://doi.org/10.1016/0092-640X(87)90013-1} {\bibfield
  {journal} {\bibinfo  {journal} {Atomic Data and Nuclear Data Tables}\
  }\textbf {\bibinfo {volume} {36}},\ \bibinfo {pages} {495 } (\bibinfo {year}
  {1987})}\BibitemShut {NoStop}%
\bibitem [{\citenamefont {Abelev}\ \emph {et~al.}(2013)\citenamefont {Abelev}
  \emph {et~al.}}]{ALICE:2013mez}%
  \BibitemOpen
  \bibfield  {author} {\bibinfo {author} {\bibfnamefont {B.}~\bibnamefont
  {Abelev}} \emph {et~al.} (\bibinfo {collaboration} {ALICE}),\ }\href
  {\doibase 10.1103/PhysRevC.88.044910} {\bibfield  {journal} {\bibinfo
  {journal} {Phys. Rev. C}\ }\textbf {\bibinfo {volume} {88}},\ \bibinfo
  {pages} {044910} (\bibinfo {year} {2013})},\ \Eprint
  {http://arxiv.org/abs/1303.0737}{arXiv:1303.0737 [hep-ex]}\BibitemShut
  {NoStop}%
\bibitem [{\citenamefont {Nemchik}\ \emph {et~al.}(2015)\citenamefont
  {Nemchik}, \citenamefont {Pasechnik},\ and\ \citenamefont
  {Potashnikova}}]{Nemchik:2014gka}%
  \BibitemOpen
  \bibfield  {author} {\bibinfo {author} {\bibfnamefont {J.}~\bibnamefont
  {Nemchik}}, \bibinfo {author} {\bibfnamefont {R.}~\bibnamefont {Pasechnik}},
  \ and\ \bibinfo {author} {\bibfnamefont {I.}~\bibnamefont {Potashnikova}},\
  }\href {\doibase 10.1140/epjc/s10052-015-3319-7} {\bibfield  {journal}
  {\bibinfo  {journal} {Eur. Phys. J. C}\ }\textbf {\bibinfo {volume} {75}},\
  \bibinfo {pages} {95} (\bibinfo {year} {2015})},\ \Eprint
  {http://arxiv.org/abs/1407.2781}{arXiv:1407.2781 [hep-ph]}\BibitemShut
  {NoStop}%
\bibitem [{\citenamefont {Voloshin}\ and\ \citenamefont
  {Poskanzer}(2000)}]{Voloshin:1999gs}%
  \BibitemOpen
  \bibfield  {author} {\bibinfo {author} {\bibfnamefont {S.~A.}\ \bibnamefont
  {Voloshin}}\ and\ \bibinfo {author} {\bibfnamefont {A.~M.}\ \bibnamefont
  {Poskanzer}},\ }\href {\doibase 10.1016/S0370-2693(00)00017-4} {\bibfield
  {journal} {\bibinfo  {journal} {Phys. Lett. B}\ }\textbf {\bibinfo {volume}
  {474}},\ \bibinfo {pages} {27} (\bibinfo {year} {2000})},\ \Eprint
  {http://arxiv.org/abs/nucl-th/9906075}{arXiv:nucl-th/9906075}\BibitemShut
  {NoStop}%
\bibitem [{\citenamefont {Andr\'es}\ \emph {et~al.}(2015)\citenamefont
  {Andr\'es}, \citenamefont {Dias~de Deus}, \citenamefont {Moscoso},
  \citenamefont {Pajares},\ and\ \citenamefont {Salgado}}]{Andres:2014xka}%
  \BibitemOpen
  \bibfield  {author} {\bibinfo {author} {\bibfnamefont {C.}~\bibnamefont
  {Andr\'es}}, \bibinfo {author} {\bibfnamefont {J.}~\bibnamefont {Dias~de
  Deus}}, \bibinfo {author} {\bibfnamefont {A.}~\bibnamefont {Moscoso}},
  \bibinfo {author} {\bibfnamefont {C.}~\bibnamefont {Pajares}}, \ and\
  \bibinfo {author} {\bibfnamefont {C.~A.}\ \bibnamefont {Salgado}},\ }\href
  {\doibase 10.1103/PhysRevC.92.034901} {\bibfield  {journal} {\bibinfo
  {journal} {Phys. Rev. C}\ }\textbf {\bibinfo {volume} {92}},\ \bibinfo
  {pages} {034901} (\bibinfo {year} {2015})},\ \Eprint
  {http://arxiv.org/abs/1405.2177}{arXiv:1405.2177 [hep-ph]}\BibitemShut
  {NoStop}%
\bibitem [{\citenamefont {Andr\'es~Casas}(2017)}]{AndresCasas:2017glq}%
  \BibitemOpen
  \bibfield  {author} {\bibinfo {author} {\bibfnamefont {C.}~\bibnamefont
  {Andr\'es~Casas}},\ }\emph {\bibinfo {title} {{Phenomenological studies of
  initial state effects and jet quenching in High-Energy Nuclear Collisions at
  LHC}}},\ \href@noop {} {Ph.D. thesis},\ \bibinfo  {school} {Santiago de
  Compostela U., IGFAE} (\bibinfo {year} {2017})\BibitemShut {NoStop}%
\bibitem [{\citenamefont {Andr\'es}\ \emph {et~al.}(2017)\citenamefont
  {Andr\'es}, \citenamefont {Braun},\ and\ \citenamefont
  {Pajares}}]{Andres:2016mla}%
  \BibitemOpen
  \bibfield  {author} {\bibinfo {author} {\bibfnamefont {C.}~\bibnamefont
  {Andr\'es}}, \bibinfo {author} {\bibfnamefont {M.}~\bibnamefont {Braun}}, \
  and\ \bibinfo {author} {\bibfnamefont {C.}~\bibnamefont {Pajares}},\ }\href
  {\doibase 10.1140/epja/i2017-12226-5} {\bibfield  {journal} {\bibinfo
  {journal} {Eur. Phys. J. A}\ }\textbf {\bibinfo {volume} {53}},\ \bibinfo
  {pages} {41} (\bibinfo {year} {2017})},\ \Eprint
  {http://arxiv.org/abs/1609.03927}{arXiv:1609.03927 [hep-ph]}\BibitemShut
  {NoStop}%
\end{thebibliography}%

\end{document}